[1994/12/01]
\documentclass[mn,fleqn]{w-art}
\usepackage{times}
\usepackage{amsmath}
\usepackage{amssymb} 
\usepackage{w-thm}
\usepackage{bm}
\usepackage{enumerate}
\usepackage[]{graphicx}
\usepackage{youngtab}

\numberwithin{equation}{section}

\chardef\bslash=`\\ 

\newcommand{\delete}{{\mathtt{delete}}}
\newcommand{\calO}{{\Lambda}}
\newcommand{\calA}{{\mathcal{A}}}
\newcommand{\calL}{{\mathcal{L}}}
\newcommand{\calB}{{\mathcal{B}}}
\newcommand{\calC}{{\mathcal{C}}}

\newcommand{\calS}{{\mathcal{S}}}
\newcommand{\calT}{{\mathcal{T}}}

\newcommand{\calSC}{\calS(\calC)}
\newcommand{\calTC}{\calT(\calC)}
\newcommand{\calSBp}{\calS(\calB)^+}
\newcommand{\calSCp}{\calS(\calC)^+}

\newcommand{\calSV}{\calS(V)}
\newcommand{\calTV}{\calT(V)}
\newcommand{\calSVp}{\calS(V)^+}
\newcommand{\calTVp}{\calT(V)^+}
\newcommand{\LSV}{\calL(\calSV)}
\newcommand{\LSVV}{\calL(\calSVp,V)}
\newcommand{\LTVV}{\calL(\calTVp,V)}
\newcommand{\calSdC}{\calS_\delta(\calC)}

\newcommand{\calTdC}{\calT_\delta(\calC)}
\newcommand{\counit}{{\varepsilon}}

\newcommand{\Deltad}{{\delta}}
\newcommand{\Deltadu}{{\underline{\Deltad}}}

\newcommand{\counitd}{{\counit_\delta}}
\newcommand{\dd}{{\mathrm{d}}}
\newcommand{\ee}{{\mathrm{e}}}
\newcommand{\eei}[1]{{I}_{#1}}
\newcommand{\Id}{{\mathrm{Id}}}
\newcommand{\Tc}{T_c}

\newcommand{\tc}{t_c}
\newcommand{\bfun}{\mathbf{1}}

\newcommand{\ix}[1]{{}_{\scriptscriptstyle(#1)}}
\newcommand{\ii}[1]{{}_{\scriptstyle(#1)}}
\newcommand{\is}[1]{{}_{\scriptscriptstyle\{ #1 \}}}
\newcommand{\isu}[1]{{}_{\scriptscriptstyle\{\underline #1 \}}}
\newcommand{\iisu}[1]{{}_{{\scriptscriptstyle{\{}}
         {\scriptstyle{\underline{#1}}}{\scriptscriptstyle{\}}} }}

\newcommand{\eval}[2][\right]{\relax
  \ifx#1\right\relax \left.\fi#2#1\rvert}

\begin{document}

\Volume{XX}
\Year{2003}
\pagespan{1}{}
\Receiveddate{15 November 2003}
\Reviseddate{30 November 2003}
\Accepteddate{2 December 2003}
\Dateposted{3 December 2003}




\title[QFT meets Hopf algebra]{Quantum field theory meets Hopf algebra}


\author[Ch. Brouder]{Christian Brouder\footnote{Corresponding author:
    e-mail: {\sf christian.brouder@impmc.jussieu.fr}, 
   Phone: +00\,33\,1\,44\,27\,50\,61\, Fax:
    +00\,33\,1\,44\,27\,37\,85}\inst{1}} 
   \address[\inst{1}]{Institut de Min\'eralogie et de Physique des
      Milieux Condens\'es,
     CNRS UMR 7590, Universit\'e Pierre et Marie Curie, 
     IPGP, 140 rue de Lourmel, 75015 Paris, France}

\thanks{}

\begin{abstract}
This paper provides a primer in quantum field theory 
(QFT) based on Hopf algebra and describes new Hopf 
algebraic constructions inspired by QFT concepts.
The following QFT concepts are introduced:
chronological products, S-matrix, Feynman diagrams,
connected diagrams, Green functions, renormalization.
The use of Hopf algebra for their definition
allows for simple recursive derivations
and leads to a correspondence between Feynman diagrams
and semi-standard Young tableaux.
Reciprocally, these concepts are
used as models to derive Hopf algebraic constructions
such as a connected coregular action or a
group structure on the linear maps from $\calSV$ to $V$.
In many cases, noncommutative analogues are derived.
\end{abstract}

\maketitle


\section{Introduction}

Although Hopf algebraic concepts were used in 
quantum field theory (QFT)
as early as 1969 \cite{Ruelle}, the real boom in the
collaboration between Hopf algebra and QFT started
with the work of Connes and Kreimer in 1998 \cite{Connes},
that spurred an enthusiastic activity partly
reviewed by Figueroa and Gracia-Bondia
\cite{FigueroaGracia}.
In these works, Hopf algebraic structures were
discovered in QFT and used to reinterpret some
aspects of renormalization theory.

The aim  of the present paper is a bit different.
As a first purpose, it tries to convince the reader
that Hopf algebra is a natural language 
for QFT. For that purpose, it uses Hopf algebraic
techniques to express important concepts of QFT:
chronological products, S-matrix, Feynman diagrams,
connected diagrams, Green functions, renormalization.
The power of Hopf algebra manifests itself
through the ease and economy with which complete
proofs can be given, with full combinatorial factors.

The second purpose of this paper is to demonstrate
that QFT concepts can help designing new 
Hopf algebraic objects. As a first example, the
connected Feynman diagrams lead us to the
definition of two coproducts on $\calSC$ and
$\calTC$ (where $\calC$ is a coalgebra, $\calSC$ the symmetric
algebra over $\calC$ and $\calTC$ the tensor algebra over $\calC$). 
These two coproducts are in a comodule coalgebra relation
and enable us to define ``connected'' 
coregular actions on $\calSC$ and $\calTC$. As a second example,
the Bogoliubov approach to renormalization
leads to a group structure on
the linear maps from $\calSV$ to $V$
and from $\calTV$ to $V$, where $V$ is a
vector space. There,
the infinitesimal bialgebraic structure of
$\calTV$ plays an essential role \cite{LodayRonco4}.
As a last example we recall that
renormalization can be considered as
a functor on bialgebras~\cite{BrouderSchmitt}.

It might be useful to explain why 
Hopf algebra is so powerful to deal
with quantum field theory. 
The first reason was given long ago by Joni and Rota~\cite{Joni}: 
the coproduct splits an object into subobjects
and the product merges two objects into a new one.
These operations are fundamental in most
combinatorial problems. Therefore, Hopf algebra
is a convenient framework to deal with combinatorics
in general and with the combinatorial problems of QFT in
particular.
The second reason has to do with the fact that Hopf algebraic
techniques efficiently exploit the recursive nature
of perturbative QFT.  If we express this property in
terms of Feynman diagrams, Hopf algebra makes it
very easy to add a new vertex to a diagram and to
deduce properties of diagrams with $n$ vertices
from properties of diagrams with $n{-}1$ vertices.
Such recursive procedure can also be carried
out directly on Feynman diagrams, without using
Hopf algebra, but it is much harder and is
the source ``egregious errors by distinguished
savants'', as Wightman put it \cite{Wightman76}.
As a consequence, many textbooks give detailed
proofs for very simple diagrams and leave as an
exercise to the reader the proof of the general case.
No such thing happens with Hopf algebras:
the present paper makes clear that a proof
for one million vertices is as simple as for two
vertices. Finally, the Hopf algebraic approach
uses naturally the fact that the unrenormalized
chronological product is an associative product.
This property is usually overlooked.

The use of Hopf algebraic techniques reveals
also that many quantum field concepts can
be defined on any cocommutative coalgebra.
Sometimes, natural noncommutative analogues of the
commutative constructions of quantum field theory can 
be found.

The Hopf algebra background of this paper can
be learnt from the first chapters of any book
on the subject, but Majid's monograph \cite{Majid}
is particularly well suited.

\section{A primer in quantum field theory}
This section provides a self-contained introduction
to QFT using Hopf algebraic tools.
Some aspects of this section were
already published in a conference proceedings
\cite{BrouderGroup24}, but complete proofs are
given here for the first time. 
We deliberately avoid the delicate analytical problems
of quantum field theory.

For some well-known physicists \cite{tHooft}, Feynman diagrams are the 
essence of QFT. Indeed, Feynman diagrams contain a complete description
of perturbative QFT, which provides its most
spectacular success: the calculation
of the gyromagnetic factor of the electron \cite{Mohr,Eides}.
Therefore, this primer goes all the way to the 
derivation of Feynman diagrams. However, it is not
restricted to a particular quantum field theory but 
is valid for any cocommutative coalgebra $\calC$
over $\mathbb{C}$. By extending the coproduct and
counit of $\calC$ to the symmetric algebra $\calSC$, we equip
$\calSC$ with the structure of a commutative and cocommutative
bialgebra. Then, we twist the product of $\calSC$ using
a Laplace pairing (or coquasitriangular structure) to
define the chronological product $\circ$.
This chronological product enables us to describe the S-matrix
of the theory, that contains all the measurable quantities.
The S-matrix is then expanded over Feynman diagrams
and the Green functions are defined.

\subsection{The coalgebra $\calC$}
In the QFT of the scalar field, the counital coalgebra $\calC$ is
generated as a vector space over $\mathbb{C}$ by the symbols
$\varphi^n(x_i)$ where $n$ runs over the nonnegative integers
and $x_i$ runs over a finite number of points in $\mathbb{R}^4$. 
The choice of a finite number of points is meant to avoid analytical
problems and is consistent with the framework of perturbative
QFT. The coproduct $\Delta'$ of $\calC$ is 
\begin{eqnarray*}
\Delta' \varphi^n(x_i) &=& \sum_{j=0}^n \binom{n}{j} 
   \varphi^j(x_i) \otimes \varphi^{n-j}(x_i),
\end{eqnarray*}
the counit $\counit'$ of  $\calC$ is
$\counit'\big(\varphi^n(x_i)\big) = \delta_{n0}$.
The coalgebra $\calC$ is cocommutative (it is a direct sum
of binomial coalgebras). Moreover, $\calC$ is a pointed
coalgebra because all its simple subcoalgebras are one-dimensional
(each simple subcoalgebra is generated by a $\varphi^0(x_i)$).

This coalgebra is chosen for comparison with QFT, but the following
construction is valid for any cocommutative coalgebra.
From the coalgebra $\calC$ we now build a
commutative and cocommutative bialgebra $\calSC$.

\subsection{The bialgebra $\calSC$}\label{defcalSCsect}
The symmetric algebra $\calSC=\bigoplus_{n=0}^\infty \calS^n(\calC)$
can be equipped with the structure of a bialgebra
over $\mathbb{C}$.
The product of the bialgebra $\calSC$ is the symmetric product
(denoted by juxtaposition) and
its coproduct $\Delta$ is defined on $\calS^1(\calC)\cong\calC$
by $\Delta a = \Delta' a$ and extended to
$\calSC$ by algebra morphism: $\Delta 1=1\otimes 1$ and $\Delta (u v) =
\sum u\ix1 v\ix1 \otimes u\ix2 v\ix2$.
The elements of $\calS^n(\calC)$ are said to be of degree $n$.
The counit $\counit$ of $\calSC$ is defined to be
equal to $\counit'$ on $\calS^1(\calC)\cong\calC$ and extended to
$\calSC$ by algebra morphism:
$\counit(1)=1$ and $\counit(u v)=\counit(u)\counit(v)$.
It can be checked that $\Delta$ is coassociative
and cocommutative \cite{Takeuchi}.
Thus, $\calSC$ is a commutative and cocommutative bialgebra which
is graded as an algebra. 
In the case of the coalgebra of the scalar field, we have
\begin{eqnarray*}
\Delta \big(
\varphi^{n_1}(x_1) \dots \varphi^{n_k}(x_k) \big) &=& 
   \sum_{j_1=0}^{n_1}\dots \sum_{j_k=0}^{n_k}
  \binom{n_1}{j_1} \cdots \binom{n_k}{j_k}
  \\&&\hspace*{5mm}
   \varphi^{j_1}(x_1)\cdots\varphi^{j_k}(x_k) \otimes 
      \varphi^{n_1-j_1}(x_1)\cdots \varphi^{n_k-j_k}(x_k).
\end{eqnarray*}
The powers $\Delta^k$ of the coproduct are called 
\emph{iterated coproducts}
and are defined by $\Delta^0 =\Id$, $\Delta^1 =\Delta $ and
$\Delta^{k+1}  = (\Id^{\otimes k}\otimes\Delta)\Delta^k $.
Their action on an element $u$ of $\calSC$ is denoted by
$\Delta^k u =\sum u\ix1\otimes\dots\otimes u\ii{k+1}$.
In the case of the scalar field, we have
\begin{lemma}\label{itercop}
If $k$ is a positive integer, the $k$-th iterated coproduct of
$\varphi^n(x)$ is
\begin{eqnarray*}
\Delta^{k-1} \varphi^n(x) &=&
  \sum_{\mathbf{m}} \frac{n!}{m_1!\dots m_k!} \varphi^{m_1}(x)\otimes 
      \dots \otimes \varphi^{m_k}(x),
\end{eqnarray*}
where $\mathbf{m}=(m_1,\dots,m_k)$ runs over all $k$-tuples of
nonnegative integers $m_1,\dots,m_k$ such that $\sum_{i=1}^k m_i=n$.
\end{lemma}
\begin{proof}
For $k=1$, we have $\Delta^0=\Id$. Thus, the left hand side of
the equality is $\varphi^n(x)$. On the other hand, we have only 
one integer $m_1$ that must be equal to $n$ because of the
constraint  $\sum_{i=1}^k m_i=n$. Therefore, the right hand side
is also $\varphi^n(x)$ and the lemma is true for $k=1$. 
Assume that the lemma is true up to $\Delta^{k-1}$. From the
definition of $\Delta^k$ and the recursion hypothesis, we have
\begin{eqnarray*}
\Delta^k \varphi^n(x) &=&
  \sum_{\mathbf{m}} \sum_{i=0}^{m_k}
    \frac{n!}{m_1!\dots m_k!} \binom{m_k}{i}
    \varphi^{m_1}(x)\otimes \dots\otimes
    \varphi^{m_{k-1}}(x)\otimes 
           \varphi^{i}(x)\otimes\varphi^{m_k-i}(x).
\end{eqnarray*}
If we define the tuple
$\mathbf{m}'=(m'_1,\dots,m'_{k+1})$ with 
$m'_j=m_j$ for $j<k$, $m'_k=i$ and $m'_{k+1}=m_k-i$,
then we see that that $\mathbf{m}'$ runs over all tuples
of $k+1$ nonnegative integers such that $\sum_{j=1}^{k+1} m'_j=n$
and we can rewrite
\begin{eqnarray*}
\Delta^k \varphi^n(x) &=&
  \sum_{\mathbf{m}'}
    \frac{n!}{m'_1!\dots m'_{k+1}!} \varphi^{m'_1}(x)\otimes 
           \dots\otimes
           \varphi^{m'_{k+1}}(x),
\end{eqnarray*}
and the lemma is proved for $\Delta^k$.
\end{proof}

Now, we equip $\calSC$ with a Laplace pairing that will
be used to twist the commutative product of $\calSC$.

\subsection{Laplace pairing}
The concept of Laplace pairing was introduced by
Rota and collaborators \cite{DRS,Grosshans}. In
the quantum group literature, it is called 
a coquasitriangular structure \cite{Majid}.
\begin{definition}
A Laplace pairing 
is a linear map $(|)$ from $\calSC\otimes \calSC$
to the complex numbers such that $(1|u)=(u|1)=\counit(u)$,
$(u v|w)=\sum (u|w\ix1)(v|w\ix2)$ and $(u|v w)=\sum (u\ix1|v)(u\ix2|w)$
for any $u$, $v$ and $w$ in $\calSC$.  
\end{definition}

The Laplace pairing of products of elements of $\calSC$
is calculated with the following lemma
\begin{lemma}\label{u1uk}
For $u^i$ and $v^j$ in $\calSC$, we have
\begin{eqnarray*}
(u^1\cdots u^k|v^1\cdots v^l)
&=&
\sum \prod_{i=1}^k\prod_{j=1}^l (u\ii{j}^i|v\ii{i}^j),
\end{eqnarray*}
where $u\ii{j}^i$ is the term in position $j$ of the
iterated coproduct $\Delta^{l-1} u^i$ and
$v\ii{i}^j$ is the term in position $i$
of the iterated coproduct $\Delta^{k-1} v^j$.
\end{lemma}
For example
$(u v w|s t)=\sum
(u\ix1|s\ix1)(u\ix2|t\ix1)(v\ix1|s\ix2)(v\ix2|t\ix2)
        (w\ix1|s\ix3)(w\ix2|t\ix3)$.
\begin{proof}
The proof is recursive. If $k=l=1$, we have
$\Delta^0 u^1=u^1$ and 
$\Delta^0 v^1=v^1$, so that lemma \ref{u1uk} becomes
$(u^1|v^1)=(u^1|v^1)$, which is true.
Now assume that the lemma is true up to $k$ and $l$
and write $u^k=st$ with $s$ and $t$ in $\calSC$.
Lemma \ref{u1uk} becomes
\begin{eqnarray}
(u^1\cdots u^k|v^1\cdots v^l)
&=&
\sum \prod_{i=1}^{k-1}\prod_{j=1}^l (u\ii{j}^i|v\ii{i}^j)
\prod_{j=1}^l ((st)\ii{j}|v\ii{k}^j).
\label{u1ukproof}
\end{eqnarray}
By algebra morphism $(st)\ii{j}=s\ii{j}t\ii{j}$ and
by the definition of the Laplace pairing and by the
coassociativity of the coproduct
$(s\ii{j}t\ii{j}|v\ii{k}^j)=
\sum(s\ii{j}|v\ii{k}^j)(t\ii{j}|v\ii{k+1}^j)$.
If we introduce this equation in (\ref{u1ukproof}) and
redefine $u^k=s$ and $u^{k+1}=t$, we obtain
Lemma \ref{u1uk} for $k+1$ and $l$. If we apply the same
reasoning to $v^l=st$, we see that the lemma is true for
$k$ and $l+1$. Thus, it is true for all $k$ and $l$.
\end{proof}

If we write lemma \ref{u1uk} with all $u^i$ and $v^j$ in $\calC$,
we see that the Laplace pairing is
entirely determined by its value on $\calC$. In other words,
once we know $(a|b)$ for all $a$ and $b$ in $\calC$,
lemma \ref{u1uk} enables us to calculate the Laplace
pairing on $\calSC$.
In the case of the algebra of the scalar field, we can
use an additional structure to
determine the usual QFT expression for $(\varphi^n(x)|\varphi^m(y))$.
\begin{lemma}\label{phinphim}
For nonnegative integers $n$ and $m$,
\begin{eqnarray*}
(\varphi^n(x)|\varphi^m(y)) &=&
\delta_{nm} n! g(x,y)^n,  
\end{eqnarray*}
where $g(x,y)=(\varphi(x)|\varphi(y))$.
\end{lemma}
\begin{proof}
The coalgebra $\calC$ was not supposed to be a bialgebra.
However, the algebra of the scalar field can be equipped with
the structure of a bialgebra. This additional structure
will be described in section \ref{simpmodsect}.
At this point, we only need the obvious product structure
$\varphi^n(x)\cdot\varphi^m(x)=\varphi^{n+m}(x)$. We
do not need to know what is the product of fields at different points
and the product is only used here as a heuristic to 
determine $(\varphi^n(x)|\varphi^m(y))$. 
We consider that the Laplace pairing satisfies its
defining properties for the product $\cdot$ of $\calC$:
$(a\cdot b|c)=\sum(a|c\ix1)(b|c\ix2)$ and
$(a|b\cdot c)=\sum(a\ix1|b)(a\ix2|c)$ for $a$, $b$ and $c$
in $\calC$.
Within this point of view, $\varphi^0(x)$ is a sort of unit
at point $x$ and 
$(\varphi^0(x)|\varphi^n(y))=(\varphi^n(x)|\varphi^0(y))=\delta_{0,n}$.

The lemma is clearly true for $n=m=1$. Assume
that it is true up to $n$ and $m$ and calculate
$(\varphi^{n+1}(x)|\varphi^m(y))=(\varphi(x)\cdot\varphi^n(x)|\varphi^m(y))$.
From the definition of a Laplace pairing, we have
\begin{eqnarray*}
(\varphi(x)\cdot\varphi^n(x)|\varphi^m(y)) &=&
\sum_{j=0}^m \binom{m}{j}
(\varphi(x)|\varphi^j(y)) (\varphi^n(x)|\varphi^{m-j}(y)).
\end{eqnarray*}
The recursion hypothesis gives us $j=1$ and $n=m-1$, so that
\begin{eqnarray*}
(\varphi^{n+1}(x)|\varphi^m(y)) &=&
m n!\delta_{n,m-1}
(\varphi(x)|\varphi(y)) (\varphi^n(x)|\varphi^{n}(y))
=\delta_{m,n+1} (n+1)! (\varphi(x)|\varphi(y))^{n+1},
\end{eqnarray*}
and the lemma is proved for $n+1$. The same reasoning leads to
the lemma for $m+1$.
\end{proof}

In QFT, the function $g(x,y)$ is a distribution \cite{Itzykson}.
Two distributions are commonly used: the Wightman function
and the Feynman propagator. The product
$g(x,y)^n$ is well-defined for Wightman functions but not
for Feynman propagators \cite{ReedSimonII}.
The solution of this problem is the first step of the
renormalization theory. In the following, we
assume that $g(x,y)$ was regularized to make it a smooth
function, so that $g(x,y)^n$ is well defined.

\subsection{Twisted product}
The Laplace pairing induces a twisted product $\circ$ on $\calSC$.
\begin{definition}
If $u$ and $v$ are elements of $\calSC$, the twisted
product of $u$ and $v$ is denoted by $u\circ v$ and
defined by $u\circ v=\sum (u\ix1|v\ix1) u\ix2 v\ix2$. 
\end{definition}
This product was introduced by Sweedler \cite{Sweedler}
as a crossed product in
Hopf algebra cohomology theory because a Laplace pairing is
a 2-cocycle. It was defined independently by Rota and Stein
as a \emph{circle product} \cite{RotaStein94}.
To become familiar with this twisted product, we first
prove a useful relation
\begin{lemma}\label{counituov}
For $u$ and $v$ in $\calSC$, we have
\begin{eqnarray*}
\counit(u\circ v) &=& (u|v)
\end{eqnarray*}
\end{lemma}
\begin{proof}
The proof is straightforward.
By linearity and algebra morphism property of the counit
\begin{eqnarray*}
\counit(u\circ v) &=& 
\sum (u\ix1|v\ix1) \counit(u\ix2 v\ix2)=
\sum (u\ix1|v\ix1) \counit(u\ix2)\counit(v\ix2).
\end{eqnarray*}
Now, by linearity of the Laplace pairing and the definition of
the counit
\begin{eqnarray*}
\counit(u\circ v) &=& 
\big(\sum u\ix1 \counit(u\ix2) | \sum v\ix1 \counit(v\ix2)\big)
=(u|v).
\end{eqnarray*}
\end{proof}

For completeness, we now prove the classical
\begin{proposition}
The twisted product $\circ$ endows $\calSC$ with the structure
of an associative and unital algebra with unit 1.
\end{proposition}
The proof is the consequence of several lemmas.
The first lemma is
\begin{lemma}\label{Deltauov}
For $u$ and $v$ in $\calSC$,
\begin{eqnarray*}
\Delta (u\circ v) &=& \sum u\ix1\circ v\ix1 \otimes  u\ix2 v\ix2.
\end{eqnarray*}
\end{lemma}
\begin{proof}
By the definition of the twisted product,
\begin{eqnarray*}
\Delta (u\circ v) &=& \sum (u\ix1|v\ix1) \Delta(u\ix2 v\ix2)
= \sum (u\ix1|v\ix1) u\ix2 v\ix2 \otimes u\ix3 v\ix3
=\sum u\ix1\circ v\ix1 \otimes  u\ix2 v\ix2,
\end{eqnarray*}
where we used the coassociativity of the coproduct.
\end{proof}
The second lemma is
\begin{lemma}\label{(u|vow)}
For $u$, $v$ and $w$ in $\calSC$,
\begin{eqnarray*}
(u|v\circ w) &=& (u\circ v|w).
\end{eqnarray*}
\end{lemma}
\begin{proof}
From the definitions of the twisted product and of the
Laplace pairing we find
\begin{eqnarray*}
(u|v\circ w) &=& \sum (v\ix1|w\ix1) (u|v\ix2 w\ix2)
= \sum (v\ix1|w\ix1) (u\ix1|v\ix2)(u\ix2| w\ix2)
\\&=&
 \sum (v\ix1|w\ix2) (u\ix2|v\ix2)(u\ix1| w\ix1),
\end{eqnarray*}
where we used the cocommutativity of the coproduct
of $u$ and $w$.
The definition of the Laplace pairing gives now
\begin{eqnarray*}
(u|v\circ w) &=& 
\sum (u\ix1 v\ix1|w) (u\ix2|v\ix2)=(u\circ v|w).
\end{eqnarray*}
\end{proof}
These two lemmas enable us to prove the associativity of the
twisted product as follows
\begin{eqnarray*}
u\circ(v\circ w) &=& \sum (u\ix1|(v\circ w)\ix1) u\ix2 (v\circ w)\ix2
=\sum (u\ix1|v\ix1\circ w\ix1) u\ix2 v\ix2 w\ix2,
\end{eqnarray*}
by lemma \ref{Deltauov}. Lemma \ref{(u|vow)} is used to
transform $(u\ix1|v\ix1\circ w\ix1)$ into
$(u\ix1\circ v\ix1|w\ix1)$, so that
\begin{eqnarray*}
u\circ(v\circ w) &=& 
\sum (u\ix1\circ v\ix1| w\ix1) u\ix2 v\ix2 w\ix2
=\sum ((u\circ v)\ix1| w\ix1) (u\circ v)\ix2 w\ix2
=(u\circ v)\circ w,
\end{eqnarray*}
where we used again lemma \ref{Deltauov} and the definition
of the twisted product. Finally, the fact that 1 is the unit
of the twisted product follows from the condition 
$(u|1)=\counit(u)$ by
\begin{eqnarray*}
u\circ 1 &=& \sum (u\ix1|1) u\ix1 = \sum \counit(u\ix1)u\ix2=u.
\end{eqnarray*}

\subsection{Iterated twisted products}
In quantum field theory, we start from an element $a$ of $\calC$,
called the Lagrangian,
and the S-matrix is defined (in the sense of formal power series
in the complex number $\lambda$)
as
\begin{eqnarray}
S &=& \exp_\circ{\lambda a} = 1 + \lambda a 
  + \frac{\lambda^2}{2!} a \circ a
  + \frac{\lambda^3}{6!} a \circ a \circ a + \dots
\end{eqnarray}
To compare this expression with that given in QFT textbooks
\cite{Itzykson,Peskin}, take $\lambda=i$.
Therefore, it is important to investigate iterated twisted
products. We shall see that Feynman diagrams arise from these
iterated products.
The main properties of iterated twisted product are consequences
of the following three lemmas
\begin{lemma}\label{Deltaa1circak}
For $u^1,\dots,u^k$ in $\calSC$ we have
\begin{eqnarray*}
\Delta (u^1\circ\dots\circ u^k) &=&
\sum u\ix1^1\circ\dots\circ u\ix1^k \otimes u\ix2^1\cdots u\ix2^k.
\end{eqnarray*}
\end{lemma}
\begin{proof}
The lemma is true for $k=2$ by lemma \ref{Deltauov}. Assume
that it is true up to $k$ and put $u^k=v\circ w$.
Then, lemma \ref{Deltauov} and the associativity of the twisted product
imply that lemma \ref{Deltaa1circak} is true for $k+1$.
\end{proof}
The next lemma gives an explicit expression for
$\counit(u^1\circ\dots\circ u^k)$
\begin{lemma}\label{epsa1circakr}
For $u^1,\dots,u^k$ in $\calSC$ we have
\begin{eqnarray*}
\counit(u^1\circ\dots\circ u^k)
&=& \sum\prod_{i=1}^{k-1}\prod_{j=i+1}^{k}
  (u\ii{j-1}^i|u\ii{i}^{j}),
\end{eqnarray*}
where $u\ii{i}^{j}$ is the term in position $i$
of the iterated coproduct 
$\Delta^{k-2} u^j$.
\end{lemma}
For example,
$\counit(u\circ v)=(u|v)$ and
$\counit(u\circ v \circ w)=\sum (u\ix1|v\ix1)(u\ix2|w\ix1)(v\ix2|w\ix2)$.
In general, 
$\counit(u^1\circ\dots\circ u^k)$ is a sum of products of
$k(k-1)/2$ Laplace pairings.
\begin{proof}
For $k=2$, lemma \ref{epsa1circakr} is true because
of lemma \ref{counituov}. Assume that it is true
up to $k$ and write $U=u^1\circ\dots\circ u^k$.
From lemma \ref{counituov} and
$U=\sum \counit(U\ix1)U\ix2$ we find
\begin{eqnarray*}
\counit(U\circ u^{k+1}) &=& (U|u^{k+1})=\sum \counit(U\ix1) (U\ix2| u^{k+1})
=
\sum \counit(u\ix1^1\circ\dots\circ u\ix1^k) 
  (u\ix2^1\cdots u\ix2^k| u^{k+1})
\\&=&
\sum \counit(u\ix1^1\circ\dots\circ u\ix1^k) 
   \prod_{n=1}^k (u\ix2^n|u\ii{n}^{k+1}),
\end{eqnarray*}
where we used lemmas \ref{Deltaa1circak} and \ref{u1uk}.
By the recursion hypothesis, we have
\begin{eqnarray*}
\counit(u^1\circ\dots \circ u^{k+1}) &=& 
\sum\prod_{i=1}^{k-1}\prod_{j=i+1}^{k}
  (u\ii{j-1}^i|u\ii{i}^{j})
   \prod_{n=1}^k (u\ii{k}^n|u\ii{n}^{k+1})
\\&=&
\sum\prod_{i=1}^{k-1}\prod_{j=i+1}^{k+1}
  (u\ii{j-1}^i|u\ii{i}^{j})
   (u\ii{k}^k|u\ii{k}^{k+1})
=
\sum\prod_{i=1}^{k}\prod_{j=i+1}^{k+1}
  (u\ii{j-1}^i|u\ii{i}^{j})
\end{eqnarray*}
and the identity is proved for the twisted product of $k+1$ elements.
\end{proof}

The last lemma completes the calculation of $u^1\circ\dots\circ u^k$
by expressing it as a linear combination of elements of $\calSC$.
\begin{lemma}\label{a1circak}
For $u^1,\dots,u^k$ in $\calSC$ we have
\begin{eqnarray*}
u^1\circ\dots\circ u^k &=&
\sum \counit(u\ix1^1\circ\dots\circ u\ix1^k) u\ix2^1\cdots u\ix2^k.
\end{eqnarray*}
\end{lemma}
\begin{proof}
To show lemma \ref{a1circak} recursively, we observe that it is true
for $k=1$ by the definition of the counit.
We assume that the property is true up to $k$ and we
define $U=u^1\circ\dots\circ u^k$.  Since, by definition,
$U\circ v=\sum (U\ix1|v\ix1) U\ix2 v\ix2$,
lemma \ref{Deltaa1circak} yields
$U\circ v=\sum (u\ix1^1\circ\dots\circ u\ix1^k|v\ix1)
  u\ix2^1\cdots u\ix2^k v\ix2$
and the result follows for the twisted product of
$k+1$ terms because of lemma \ref{counituov}.
\end{proof}

\subsection{Application to the scalar field}
\label{scalfieldsect}
If we apply these results to the coalgebra of 
the scalar field, we obtain the following expression
for the iterated twisted product
\begin{proposition}\label{iterophi}
\begin{eqnarray}
\phi^{n_1}(x_1)\circ\dots \circ \phi^{n_k}(x_k) &=&
\sum_{p_1=0}^{n_1}\cdots\sum_{p_k=0}^{n_k}
\binom{n_1}{p_1}\dots \binom{n_k}{p_k}
\counit\big(\phi^{p_1}(x_1)\circ\dots \circ \phi^{p_k}(x_k)\big)
\nonumber\\&&\hspace*{10mm}
\phi^{n_1-p_1}(x_1)\dots \phi^{n_k-p_k}(x_k),
\label{EG}
\end{eqnarray}
with
\begin{eqnarray}
\counit\big(\phi^{p_1}(x_1)\circ\dots \circ \phi^{p_k}(x_k)\big)
 &=& p_1!\dots p_k!
\sum_{M} \prod_{i=1}^{k-1}\prod_{j=i+1}^k 
    \frac{g(x_i,x_j)^{m_{ij}}}{m_{ij}!},
\label{omegab}
\end{eqnarray}
where the sum is over all symmetric $k\times k$ matrices $M$ of
nonnegative
integers $m_{ij}$ such that $\sum_{j=1}^k m_{ij}=p_i$ with
$m_{ii}=0$ for all $i$.
\end{proposition}
\begin{proof}
Equation \eqref{EG} is a simple rewriting of lemma
\ref{a1circak} for 
$u^1=\phi^{(n_1)}(x_1)$,\dots,$u^k=\phi^{(n_k)}(x_k)$. For the proof
of \eqref{omegab}, we first recall from lemma \ref{itercop}
that
\begin{eqnarray*}
\Delta^{k-2} \phi^{p_i}(x_i) & = & \sum \frac{p_i!}{q_{i1}!\dots q_{ik-1}!}
   \phi^{q_{i1}}(x_i)\otimes\dots\otimes \phi^{q_{ik-1}}(x_i),
\end{eqnarray*}
where the sum is over all nonnegative integers $q_{ij}$ such that
$\sum_{j=1}^{k-1} q_{ij}=p_i$. Thus, lemma
\ref{epsa1circakr} becomes
\begin{eqnarray*}
\counit\big(\phi^{p_1}(x_1)\circ\dots \circ \phi^{p_k}(x_k)\big) &=& \sum_Q 
     \Big(\prod_{i=1}^k \frac{p_i!}{\prod_{j=1}^{k-1} q_{ij}!}\Big)
     \prod_{i=1}^{k-1}\prod_{j=i+1}^k 
      \big( \varphi^{q_{ij-1}}(x_i)|\varphi^{q_{ji}}(x_j)\big),
\end{eqnarray*}
where the sum is over all $k\times (k-1)$ matrices $Q$ with matrix elements
$q_{ij}$ such that $\sum_{j=1}^{k-1} q_{ij}=p_i$.
Lemma \ref{phinphim} implies that $q_{ij-1}=q_{ji}$. 
It is very convenient to associate to each matrix $Q$
a $k\times k$ matrix $M$ with matrix elements
$m_{ji}=q_{ji}$ if $i<j$, $m_{ii}=0$ and
$m_{ji}=q_{ji-1}$ if $j<i$. With this definition,
the condition $q_{ij-1}=q_{ji}$ implies that the
matrix $M$ is symmetric.
The condition $\sum_{j=1}^{k-1} q_{ij}=p_i$
implies $\sum_{j=1}^k m_{ji}=p_i$ and we recover
the condition of proposition \ref{iterophi}.
It remains to show that the
combinatorial factors come out correctly.
Lemma \ref{phinphim} gives us the combinatorial factor
$\prod_{i=1}^{k-1}\prod_{j=i+1}^k m_{ij}!$.
On the other hand, we have
\begin{eqnarray*}
\prod_{i=1}^k\prod_{j=1}^{k-1} q_{ij}! &=&
\Big(\prod_{i=2}^k\prod_{j=1}^{i-1} q_{ij}!\Big)
\Big(\prod_{i=1}^{k-1}\prod_{j=i}^{k-1} q_{ij}!\Big)
=
\Big(\prod_{i=2}^k\prod_{j=1}^{i-1} m_{ij}!\Big)
\Big(\prod_{i=1}^{k-1}\prod_{j=i}^{k-1} m_{ij+1}!\Big)
\\&=&
\Big(\prod_{j=1}^{k-1}\prod_{i=j+1}^{k} m_{ij}!\Big)
\Big(\prod_{i=1}^{k-1}\prod_{j=i+1}^{k} m_{ij}!\Big)
=
\Big(\prod_{i=1}^{k-1}\prod_{j=i+1}^{k} m_{ij}!\Big)^2,
\end{eqnarray*}
by symmetry of $M$. This completes the proof
\end{proof}
If $p_1$ is zero, equation \eqref{omegab} 
gives us
$\counit\big(\phi^{0}(x_1)\circ\phi^{p_2}(x_2)\dots \circ \phi^{p_k}(x_k)\big)
=\counit\big(\phi^{p_2}(x_2)\circ\dots \circ \phi^{p_k}(x_k)\big)$.
Thus, for any $u\in \calSC$, we have
$\counit\big(\phi^{0}(x)\circ u\big)=\counit(u)$.
In other words, by lemma \ref{counituov},
the definition of the Laplace pairing is
supplemented with the condition
$\big(\phi^{0}(x)| u\big)=\counit(u)$.
This is consistent with the remark made in the proof of lemma
\ref{phinphim}.
To generalize this convention
we use the fact that any cocommutative coalgebra $\calC$
over $\mathbb{C}$ is pointed (see \cite{Abe}, p.~80).
As a consequence, if $G(\calC)$ denotes the
set of group-like elements of $\calC$, then
the coradical $\calC_0$ of $\calC$
is generated by the elements of $G(\calC)$
and there is a coideal $I$ of $\calC$ such that
$\calC=\calC_0 \oplus I$ and $\counit(I)=0$. Moreover, 
$\counit(g)=1$ for $g\in G(\calC)$ 
(see \cite{Montgomery}, section 5.4 and \cite{Parker}).
Then, for any element $g\in G(\calC)$ and $u\in \calSC$, we 
define $(g|u)=\counit(u)$, so that
$\counit(g \circ u)= \counit(u)$
and $g\circ u=gu$.

\subsection{Relation with Feynman diagrams}
To clarify the relation between equation \eqref{omegab}
and Feynman diagrams, we first give a definition
of the type of Feynman diagrams we consider.
In this paper, a Feynman diagram is a vertex-labelled undirected graph
without loop. The word loop is used here in the graph-theoretical
sense of an edge that goes from a vertex to itself.
Such a loop is called a tadpole in the QFT jargon.
The vertices are labelled by spacetime points
$x_i$. If $k$ is the number of vertices of a Feynman diagram $\gamma$,
the adjacency matrix $M$ of $\gamma$ is a $k$x$k$ integer
matrix where $m_{ij}$ denotes the number of edges between
vertex $i$ and vertex $j$. The absence of loops means
that the diagonal of $M$ is zero.
The valence of a vertex is the number of edges incident to it.

The vertex-labelled Feynman diagrams that we 
use here are well known (\cite{Itzykson}, p.~265)
but not as common in the literature as
the Feynman diagrams where the edges are labelled by momenta
(\cite{Itzykson}, p.~268). However, the latter Feynman
diagrams are not as general as the vertex-labelled ones
because they assume that
the physical system is translation invariant and this is not true
in the presence of an external potential or in curved spacetime.

It is now clear that, in proposition \ref{iterophi}, each matrix $M$ 
is the adjacency matrix of a Feynman diagram $\gamma$. The value of 
$\gamma$ is the quantity
\begin{eqnarray}
U(\gamma) &=& p_1!\dots p_k!  \prod_{i=1}^{k-1}\prod_{j=i+1}^k 
    \frac{g(x_i,x_j)^{m_{ij}}}{m_{ij}!}.
\label{Ugamma}
\end{eqnarray}

To clarify this matter, it is convenient to give a few examples.
\begin{example}\label{examp33}
We consider the case of
$\counit\big(\varphi^3(x)\circ\varphi^3(y)\big)$. If we compare with
proposition \ref{iterophi}, we have $k=2$, $p_1=3$ and
$p_2=3$. The only nonnegative integer 2x2 symmetric
matrix $M$ with zero diagonal such that
$\sum_{j=1}^k m_{ij}=p_i$ is
\begin{eqnarray*}
         M = \left(\begin{array}{cc} 0 & 3 \\
                              3 & 0 \end{array}\right).
\end{eqnarray*}
Thus, according to the general formula,
$\counit\big(\varphi^3(x)\circ\varphi^3(y)\big) = 3!g(x,y)^3 $.

The matrix $M$
is the adjacency matrix of the following Feynman diagram $\gamma$, which
is called the \emph{setting sun}.
\begin{figure}[!ht]
    \includegraphics[width=4cm]{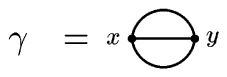}
\end{figure}
\end{example}

\begin{example}\label{examp332}
We consider $\counit\big(\varphi^3(x)\circ\varphi^3(y)\circ\varphi^2(z)\big)$. 
We have now $k=3$, $p_1=3$, $p_2=3$ and
$p_3=2$. The only matrix that satisfies the conditions is
\begin{eqnarray*}
       M = \left(\begin{array}{ccc} 0 & 2 & 1\\
                             2 & 0 & 1\\
                             1 & 1 & 0 \end{array}\right).
\end{eqnarray*}
Thus, according to the general formula,
$\counit\big(\varphi^3(x)\circ\varphi^3(y)\circ\varphi^2(z)\big) = 
  3! 3! g(x,y)^2 g(x,z) g(y,z)$.

The matrix $M$
is the adjacency matrix of the following Feynman diagram $\gamma$, which
is called the \emph{ice cream}.
\begin{figure}[!ht]
    \includegraphics[width=4cm]{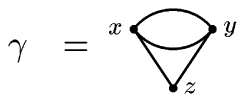}
\end{figure}
\end{example}

\begin{example}\label{examp2222}
Finally, we consider 
$\counit\big(\varphi^2(x_1)\circ\varphi^2(x_2)\circ\varphi^2(x_3)\circ\varphi^2(x_4)\big)$. 
Six matrices $M$ satisfy the conditions:
\begin{eqnarray*}
       M_1 &=& \left(\begin{array}{cccc} 0 & 0 & 0 & 2\\
                             0 & 0 & 2 & 0\\
                             0 & 2 & 0 & 0\\
                             2 & 0 & 0 & 0\end{array}\right),
       \quad
       M_2=\left(\begin{array}{cccc} 0 & 0 & 1 & 1\\
                             0 & 0 & 1 & 1\\
                             1 & 1 & 0 & 0\\
                             1 & 1 & 0 & 0\end{array}\right),
        \quad
       M_3= \left(\begin{array}{cccc} 0 & 0 & 2 & 0\\
                             0 & 0 & 0 & 2\\
                             2 & 0 & 0 & 0\\
                             0 & 2 & 0 & 0\end{array}\right),
        \\
       M_4 &=& \left(\begin{array}{cccc} 0 & 1 & 0 & 1\\
                             1 & 0 & 1 & 0\\
                             0 & 1 & 0 & 1\\
                             1 & 0 & 1 & 0\end{array}\right),
         \quad
       M_5 = \left(\begin{array}{cccc} 0 & 1 & 1 & 0\\
                             1 & 0 & 0 & 1\\
                             1 & 0 & 0 & 1\\
                              0 & 1 & 1 & 0\end{array}\right),
          \quad
       M_6 = \left(\begin{array}{cccc} 0 & 2 & 0 & 0\\
                             2 & 0 & 0 & 0\\
                             0 & 0 & 0 & 2\\
                             0 & 0 & 2 & 0\end{array}\right).
\end{eqnarray*}
The matrices $M_i$ enable us to write 
the value of
$\counit\big(\varphi^2(x_1)\circ\varphi^2(x_2)\circ\varphi^2(x_3)\circ\varphi^2(x_4)\big)$
as
\begin{eqnarray*}
&& 4 g(x_1,x_4)^2 g(x_2,x_3)^2 
+
16 g(x_1,x_3) g(x_1,x_4) g(x_2,x_3) g(x_2,x_4)
+
4 g(x_1,x_3)^2 g(x_2,x_4)^2 
\\&&
+
16 g(x_1,x_2) g(x_1,x_4) g(x_2,x_3) g(x_3,x_4)
+
16 g(x_1,x_2) g(x_1,x_3) g(x_2,x_4) g(x_3,x_4)
\\&&
+
4 g(x_1,x_2)^2 g(x_3,x_4)^2.
\end{eqnarray*}
The matrices $M_i$
are the adjacency matrices of the following Feynman diagrams $\gamma$.
\begin{figure}[!ht]
    \includegraphics[width=15cm]{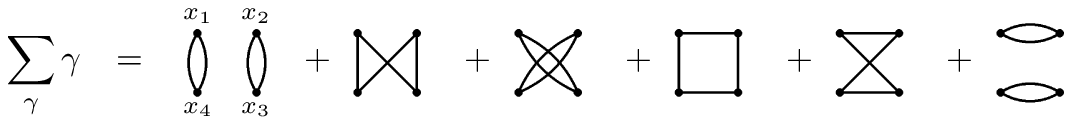}
\end{figure}
\end{example}

Note that the first, third and last Feynman diagrams 
are disconnected. This will be important in the following.
For a specific Lagrangian (for example $\varphi^n$ \cite{Itzykson}),
the value of a Feynman diagram defined in textbooks \cite{Itzykson}
is a bit different from equation \eqref{Ugamma} because an integral
over the spacetime points labelling vertices with valence $n$
is added. We do not use this convention here because we consider a 
general Lagrangian.

\subsection{Enumeration of Feynman diagrams}
Various authors
\cite{Read59,Read60,Read,Burge}
studied the number of matrices $M$ (or of Feynman diagrams)
corresponding to 
$\counit\big(\varphi^{n_1}(x_1)\circ\dots\circ\varphi^{n_k}(x_k)\big)$.
\begin{proposition}
The number of Feynman diagrams generated by
$\counit\big(\varphi^{n_1}(x_1)\circ\dots\circ\varphi^{n_k}(x_k)\big)$
is the coefficient of 
$z_1^{n_1}\dots z_k^{n_k}$
in $\prod_{i<j} (1-z_i z_j)^{-1}$.
\end{proposition}
Since we do not know of any published simple proof of this
proposition, we provide the following one.
\begin{proof}
Let $N_k(n_1,\dots,n_k)$ denote the number of 
Feynman diagrams 
of $\counit\big(\varphi^{n_1}(x_1)\circ
\dots\circ\varphi^{n_k}(x_k)\big)$
and $f_k(z_1,\dots,z_k)$ the generating function
\begin{eqnarray*}
f_k(z_1,\dots,z_k) &=&
\sum_{n_1\dots n_k} N_k(n_1,\dots,n_k) z_1^{n_1}\dots z_k^{n_k}.
\end{eqnarray*}
When $k=2$, we have
$\counit\big(\varphi^{n_1}(x_1)\circ \varphi^{n_2}(x_2)\big)=
\delta_{n_1,n_2} n_1! (\varphi(x_1)|\varphi(x_2))^{n_1}$
so there is no diagram if $n_1\not= n_2$ and 1
if $n_1=n_2$ ($n_1$ lines linking $x_1$ and $x_2$).
Therefore
\begin{eqnarray*}
f_2(z_1,z_2) &=&
\sum_{n_1,n_2} N_2(n_1,n_2) z_1^{n_1} z_2^{n_2} =
\sum_{n=0}^\infty z_1^{n} z_2^{n}
= \frac{1}{1-z_1 z_2}.
\end{eqnarray*}
Assume that you know $N_l(n_1,\dots,n_l)$ up to $l=k-1$.
To calculate it for $k$, take a matrix representing a
diagram for $k$ and call $i_1, \dots, i_{k-1}$ its last
line (recall that the diagonal is zero so that $i_{k}=0$).
The matrix obtained by removing the last line and the
last column encodes a diagram for
$\counit\big(\varphi^{n_1-i_1}(x_1)\dots
\varphi^{n_{k-1}-i_{k-1}}(x_{k-1})\big)$.
Therefore,
\begin{eqnarray*}
N_{k}(n_1,\dots,n_{k}) &=&
\sum_{i_1+\dots+i_{k-1}=n_{k}} N_{k-1}(n_1-i_1,\dots,n_{k-1}-i_{k-1}).
\end{eqnarray*}
This gives us
\begin{eqnarray*}
f_{k}(z_1,\dots,z_{k}) &=&
\sum_{n_1,\dots,n_{k}} 
N_{k}(n_1,\dots,n_{k})z_1^{n_1}\dots z_{k}^{n_{k}}
\\&=&
\sum_{n_1,\dots,n_{k}} z_1^{n_1}\dots z_{k}^{n_{k}}
\sum_{i_1+\dots+i_{k-1}=n_{k}} N_{k-1}(n_1-i_1,\dots,n_{k-1}-i_{k-1})
\\&=&
\sum_{n_{k}} z_{k}^{n_{k}}
\sum_{i_1+\dots+i_{k-1}=n_{k}} 
z_1^{i_1}\dots z_{k-1}^{i_{k-1}}
\sum_{j_1,\dots,j_{k-1}} N_{k-1}(j_1,\dots,j_{k-1})
z_1^{j_1}\dots z_{k-1}^{j_{k-1}},
\end{eqnarray*}
where we put $n_l=i_l+j_l$ for $l=1,\dots,k-1$.
The sum over $j_l$ is the generating function $f_{k-1}$ so
\begin{eqnarray*}
f_{k}(z_1,\dots,z_{k}) &=&
\sum_{n_{k}} z_{k}^{n_{k}}
\sum_{i_1+\dots+i_{k-1}=n_{k}} 
z_1^{i_1}\dots z_{k-1}^{i_{k-1}}
f_{k-1}(z_1,\dots,z_{k-1})
\\&=&
\sum_{i_1,\dots,i_{k-1}} 
(z_1 z_k)^{i_1}\dots (z_{k-1}z_k)^{i_{k-1}}
f_{k-1}(z_1,\dots,z_{k-1})
\\&=&
\frac{f_{k-1}(z_1,\dots,z_{k-1})}
{(1-z_1 z_k)\dots (1-z_{k-1}z_k)}.
\end{eqnarray*}
So finally
\begin{eqnarray*}
f_{k}(z_1,\dots,z_{k}) &=&
\prod_{j=2}^k\prod_{i=1}^{j-1} \frac{1}{1-z_i z_j}.
\end{eqnarray*}
\end{proof}
Note that, by a classical identity of the theory
of symmetric functions (see \cite{MacDonald}, p.~77),
\begin{eqnarray*}
\prod_{i<j} (1-z_iz_j)^{-1} &=& \sum_{\lambda} s_\lambda,
\end{eqnarray*}
where $s_\lambda$ is the Schur function for partition $\lambda$
and where the sum runs over all partitions $\lambda$
having an even number of parts of any given magnitude,
for example $\{1^2\}$, $\{2^2\}$, $\{1^4\}$, $\{2^2 1^2\}$,
$\{3^2\}$, $\{1^6\}$.

\subsection{Feynman diagrams and Young tableaux}
Burge \cite{Burge} proposed algorithms to generate different types
of graphs based on a correspondence with semi-standard 
Young tableaux.  We recall that a Young diagram is a collection
of boxes arranged in left-justified rows, with a weakly
decreasing number of boxes in each row \cite{FultonYoung}.
For example {\scriptsize{\yng(3,3,1,1)}} is a Young diagram.
A Young tableau is a Young diagram where each box contains a
strictly positive integer. A Young tableau is semi-standard if
the numbers contained in the boxes are weakly increasing from left to
right in each row and strictly increasing down each column.
For example, $Y$={\scriptsize{\young(113,224,3,4)}} is a semi-standard 
Young tableau.
The number of semi-standard Young tableaux of a given shape
and content is given by the Kostka numbers \cite{FultonYoung}.

To obtain all the Feynman diagrams contributing to
$t\big(\varphi^{n_1}(x_1)\dots\varphi^{n_k}(x_k)\big)$,
list all the semi-standard Young tableaux with
$n_1+\dots+n_k=2n$ boxes, where all columns of $Y$
have an even number of boxes, filled with $n_1$ times the symbol 1,
\dots, $n_k$ times the symbol $k$.  For instance, our
example tableau $Y$ corresponds to
$t\big(\varphi^{2}(x_1)\varphi^{2}(x_2)\varphi^{2}(x_3)\varphi^{2}(x_4)\big)$.
There are efficient algorithms to generate this list \cite{FackJeugt}.

Then, for each tableau $Y$ of the list, 
use the Robinson-Schensted-Knuth 
correspondence \cite{Knuth} for the pair $(Y,Y)$
to generate the Feynman diagram.
This relation between Feynman diagrams and semi-standard Young
tableaux is not widely known. Thus, it is useful
to give it here in some detail.

A box in a Young tableau is a \emph{boundary box} if 
it is the rightmost box of a row and if it has no
box below it. If a box is referenced by a pair
$(r,c)$ where $r$ is the row number and $c$ the column number
(counted from the upper-left corner), then the boundary boxes 
of our example tableau $Y$ are
the boxes $(2,3)$ and $(4,1)$ containing the
numbers 4 and 4, respectively.
We define now the algorithm $\delete$ taking a tableau $Y$ and 
a boundary box $(r,c)$ and building a tableau $Y'$
and a number $k$.
If $r=1$, then $c$ is the last column of the first row.
The result of $\delete(Y;(1,c))$ is the tableau $Y'$
obtained from the tableau $Y$ by removing box $(1,c)$
and $k$ is the number contained in box $(1,c)$ of $Y$.
If $r>1$, box $(r,c)$ is removed from $Y$ and 
the number $i$ contained in box $(r,c)$ replaces the first
number in row $r{-}1$ (from right to left)
that is strictly smaller than $i$. The number $j$
it replaces is then moved to row $r{-}2$ with the same
rule, until row 1 is reached. The tableau thus obtained
is $Y'$ and the number replaced in the first row is $k$
(see Knuth's paper \cite{Knuth} for a more
computer-friendly algorithm).

In our example $Y$, if we choose the boundary box $(2,3)$,
the number 4 of box $(2,3)$ is moved to the first row,
where it replaces the number 3, so that
$Y'$={\scriptsize{\young(114,22,3,4)}} and $k=3$.
If we choose the boundary box $(4,1)$,
we have successively
{\scriptsize{\young(113,224,4)}}, where 4 replaces a 3;
moving this 3 to the second row gives
{\scriptsize{\young(113,234,4)}}, where 3 replaces a 2;
moving this 2 to the first row gives
$Y'$={\scriptsize{\young(123,234,4)}} and the
replaced number in the first row is $k=1$.

To generate the Feynman diagram corresponding to
a given tableau $Y$ (with $2n$ boxes),
we first need to define $n$ pairs of integers
$(u_k,v_k)$, where $k=1,\dots,n$ 
(each pair represents an edge of the diagram).
Let $u_n$ be the largest number contained in the boxes of $Y$
and, among the boxes of $Y$ containing the number $u_n$,
let $(r,c)$ be the one with largest $c$.
Let $Y_1$ be the tableau obtained from $Y$ by removing box $(r,c)$.
Calculate $\delete(Y_1;(r{-}1,c))$ to obtain a tableau $Y'_1$
and a number $k$. Assign $v_n=k$ and repeat the procedure on $Y'_1$. 
This gives $n$ pairs $(u_1,v_1),\dots,(u_n,v_n)$,
with $u_n>v_n$.

Let us apply this procedure to our example tableau $Y$.
It contains 8 boxes, so that $n=4$. Its largest number
is $4$, the rightmost box containing it is $(2,3)$
and the Young tableau obtained by removing this box from $Y$ is
$Y_1$={\scriptsize{\young(113,22,3,4)}}. If we apply $\delete$
to $Y_1$ and boundary box $(1,3)$ we obtain
$Y'_1$={\scriptsize{\young(11,22,3,4)}} and $k=3$. Thus,
$(u_4,v_4)=(4,3)$. 
We have now $u_3=4$ and the box $(4,1)$. We apply the
same procedure to get
$Y'_2$={\scriptsize{\young(12,23)}} and $k=1$, so that $(u_3,v_3)=(4,1)$.
Doing this again gives us
$Y'_3$={\scriptsize{\young(1,2)}} and $(u_2,v_2)=(3,1)$.
Finally $(u_1,v_1)=(2,1)$.

Then, the Feynman diagram is generated as follows:
let $k$ be the largest number contained in $Y$,
draw $k$ points labelled $x_1,\dots,x_k$.
Then, for each pair $(u_i,v_i)$, draw an edge
between $x_a$ and $x_b$, where $a=u_i$ and $b=v_i$.
Note that our example tableau $Y$ corresponds
to the fourth diagram of example \ref{examp2222}.

Similarly, the adjacency matrix $M$ is built as follows:
let $k$ be the largest number contained in $Y$
and $M$ the $k\times k$ null matrix.
For $i=1$ to $n$ increase
$m_{ab}$ and $m_{ba}$ by one, where $a=u_i$ and $b=v_i$.

The semi-standard Young tableaux corresponding to the diagrams of
examples \ref{examp33}, \ref{examp332}
and \ref{examp2222} are, respectively,
{\scriptsize{\young(111,222)}},
{\scriptsize{\young(1112,2233)}} and
{\scriptsize{\young(11,22,33,44),
\young(112,234,3,4),
\young(1122,3344),
\young(113,224,3,4),
\young(1123,2344),
\young(1133,2244)}}.

\subsection{Chronological product and Green functions}
\label{chronsect}
If the Laplace coupling is symmetric, i.e. if
$(u|v)=(v|u)$ for all $u$ and $v$ in $\calSC$, then
the twisted product is commutative and we can define
a map $T$ from $\calSC$ to $\calSC$ by
$T(uv)=T(u)\circ T(v)$ and
$T(a)=a$ for $a\in\calC$.
In particular, if $a_1,\dots,a_k$ are elements of $\calC$,
then $T(a_1\dots a_k)=a_1\circ\dots\circ a_k$.
Using lemma \ref{a1circak}, we see that
$T(u)=\sum t(u\ix1) u\ix2$, where $t$ is a linear map
from $\calSC$ to $\mathbb{C}$ defined by
$t(u)=\counit\big(T(u)\big)$.
For historical reasons, the map $T$ is
called the \emph{chronological product}
or the \emph{time-ordered product}
\cite{Dyson,Itzykson}.
The map $t$ is an element of $\calSC^*$,
the dual of $\calSC$. In that sense, the
chronological product is the right coregular action
of $\calSC^*$ on $\calSC$ by
$T(u)=u\triangleleft t = \sum t(u\ix1)u\ix2$
and $\calSC$ is a right $\calSC^*$-module
if $\calSC^*$ is endowed with the convolution product
(see \cite{Majid} p.~21).
The Hopf algebraic properties of $T$ were discussed
in detail in \cite{BrouderQG}.

If we take the convention that $(g|u)=\counit(u)$
for any element $g\in G(\calC)$ and $u\in \calSC$
(see the end of section \ref{scalfieldsect}), then
$T(gu)=g T(u)$ and $t(gu)=t(u)$. In particular, for the
algebra of fields, $T(\varphi^0(x)u)=\varphi^0(x) T(u)$ and
$t(\varphi^0(x)u)=t(u)$.

The map $T$ enables us to write the S-matrix as
$S=T\big(\exp(\lambda a)\big)$, where
$\exp(\lambda a)=1+\sum_{n=1}^\infty (\lambda^n/n!) a^n$,
where the product $a^n$ is the product in $\calSC$.
Moreover, the map $t$ is useful to define the
Green functions of a theory:
\begin{definition}\label{defG}
The $n$-point Green function of the scalar field theory 
with Lagrangian $a$ is the function
\begin{eqnarray*}
G(x_1,\dots,x_k) &=& 
t\big(\varphi(x_1)\dots\varphi(x_k) \exp(\lambda a)\big).
\end{eqnarray*}
\end{definition}
In the usual definition of the Green function, the
right hand side of the previous equation is divided by
$t\big(\exp(\lambda a)\big)$, because of  
the Gell-Mann and Low theorem \cite{GellMann}.
We omit this denominator for notational convenience.
The usefulness of the Green functions stems from the fact that
most important physical quantities can be expressed in terms
of Green functions with a small number of arguments. 
For instance, the charge density
is proportional to $G(x,x)$ \cite{Fetter}, the optical spectrum 
is a linear function of $G(x_1,x_2,x_3,x_4)$ \cite{Strinati}, etc.

Let us consider the example of the $\varphi^4$ theory
with Lagrangian $a=\int_{\mathbb{R}^4} \varphi^4(x) \dd x$.
Strictly speaking, we are not allowed to consider an 
infinite number of points as in the integral over
$\mathbb{R}^4$, but this drawback can be cured in the
perturbative regime, where $ \exp(\lambda a)$
is expanded as a series in $\lambda$ and each 
term $a^n$ is rewritten
\begin{eqnarray*}
a^n &=& \int_{\mathbb{R}^{4n}} \dd y_1 \dots \dd y_n
         \varphi^4(y_1) \dots \varphi^4(y_n).
\end{eqnarray*}
Thus, the perturbative expansion of the Green function is
\begin{eqnarray*}
G(x_1,\dots,x_k) &=& \sum_{n=0}^\infty \frac{\lambda^n}{n!}
\int_{\mathbb{R}^{4n}} \dd y_1 \dots \dd y_n
t\big(\varphi(x_1)\dots\varphi(x_k) \varphi^4(y_1) \dots \varphi^4(y_n)\big).
\end{eqnarray*}
Then, in each term of the expansion, the
coregular action $t$ acts on a finite number of points
and the present formulation is valid 
(up to renormalization).
For instance, the first nonzero terms of $G(x_1,x_2)$ are
\begin{eqnarray*}
G(x_1,x_2) &=& 
    t\big(\varphi(x_1)\varphi(x_2)\big)
+ \frac{\lambda^2}{2} 
\int_{\mathbb{R}^{8}} \dd y_1 \dd y_2
t\big(\varphi(x_1)\varphi(x_2) \varphi^4(y_1) \varphi^4(y_2)\big) 
+ O(\lambda^3),
\\&=&
g(x_1,x_2) + \lambda^2 \int_{\mathbb{R}^{8}} \dd y_1 \dd y_2
\Big( 48 g(x_1,y_1) g(y_1,y_2)^3 g(y_2,x_2)
\\&&
     + 48 g(x_1,y_2) g(y_1,y_2)^3 g(y_1,x_2)
     + 12 g(x_1,x_2) g(y_1,y_2)^4\Big) + O(\lambda^3)
\end{eqnarray*}
In terms of Feynman diagrams, this gives us
\begin{figure}[!ht]
    \includegraphics[width=12cm]{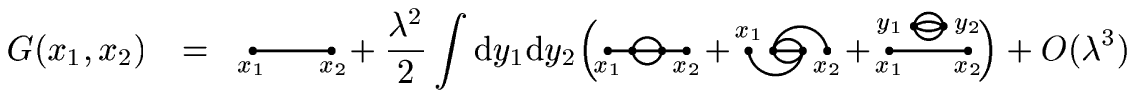}
\end{figure}

For the clarity of the figure, the labels $y_1$ and $y_2$ were given
explicitly only for the last diagram.
The three diagrams under the integral sign correspond to the
three matrices
\begin{eqnarray*}
       M_1 &=& \left(\begin{array}{cccc} 0 & 0 & 1 & 0\\
                             0 & 0 & 0 & 1\\
                             1 & 0 & 0 & 3\\
                             0 & 1 & 3 & 0\end{array}\right),
       \quad
       M_2=\left(\begin{array}{cccc} 0 & 0 & 0 & 1\\
                             0 & 0 & 1 & 0\\
                             0 & 1 & 0 & 3\\
                             1 & 0 & 3 & 0\end{array}\right),
        \quad
       M_3= \left(\begin{array}{cccc} 0 & 1 & 0 & 0\\
                             1 & 0 & 0 & 0\\
                             0 & 0 & 0 & 4\\
                             0 & 0 & 4 & 0\end{array}\right).
\end{eqnarray*}
Note that the first and second integrals
have the same value because $g(y_1,y_2)=g(y_2,y_1)$. 
Their sum would be counted as a single
diagram in the standard theory. The advantage of the present
procedure is that we do not have to determine the number of
elements of the symmetry group of the diagrams.
Note also that the last term is not connected.
If we had used the standard definition
of Green functions, the disconnected graph
(i.e. the third term of the integral)
would be cancelled by the denominator
$t\big(\exp(\lambda a)\big)$.

We considered only Feynman diagrams without tadpoles
(a tadpole is an edge going from a vertex to itself).
For some applications, it is useful to consider diagrams
that can possibly contain tadpoles. Burge defined a similar correspondence
for this more general case \cite{Burge}. The semi-standard
tableaux are now required to have an even number of boxes in
each row (and no condition on the columns). 
In the corresponding adjacency matrix, the
tadpoles attached to vertex $x_k$ are described by 
assigning to the diagonal matrix element $m_{kk}$
the value of twice the number of tadpoles attached to
$x_k$.

\subsection{The noncocommutative case}
When the coalgebra $\calC$ is not cocommutative, the 
above construction is not valid because the twisted
product is not associative. However, it is still
possible to define a bialgebra $\calTC$ from any
coalgebra $\calC$ over $\mathbb{C}$ as follows.
As an algebra, $\calTC$ is the tensor algebra of
$\calC$: $\calTC=\bigoplus_{n=0}^\infty \calT^n(\calC)$.
The coproduct $\Delta$ of $\calTC$ is defined by extending
the coproduct of $\calC$ by algebra morphism
and the counit $\counit$ of $\calTC$ is defined by extending
the counit of $\calC$ by algebra morphism.
With this coproduct and counit, the algebra $\calTC$
becomes a bialgebra.
This bialgebra was used by Ritter \cite{Ritter} in the
framework of noncommutative QFT.

For any element $t$ of $\calTC^*$,
the dual of $\calTC$, we can define the
right coregular action
of $\calTC^*$ on $\calTC$ by
$T(u)=u\triangleleft t = \sum t(u\ix1)u\ix2$.
If $\calTC^*$ is endowed with the convolution product
$(t * t')(u)=\sum t(u\ix1) t'(u\ix2)$, then
$\calTC^*$ is a unital associative algebra with unit $\counit$.
Moreover,
\begin{lemma}
$\calTC$ is a right $\calTC^*$-module, i.e.
$u\triangleleft \counit=u$ and
$(u\triangleleft t)\triangleleft t'=u\triangleleft (t *t')$.
\end{lemma}
\begin{proof}
By definition of the action and of the counit, we have
$u\triangleleft \counit=\sum \counit(u\ix1) u\ix2=u$.
By definition of the action,
\begin{eqnarray*}
(u\triangleleft t)\triangleleft t' &=& 
\sum \big(t(u\ix1) u\ix2\big)\triangleleft t' = 
\sum t(u\ix1) (u\ix2\triangleleft t') = 
\sum t(u\ix1) t'(u\ix2) u\ix3
\\&=& 
 \sum (t * t')(u\ix1) u\ix2 = u\triangleleft (t *t').
\end{eqnarray*}
\end{proof}

\section{Connected chronological products}
In example \ref{examp2222}, we saw that some of the
Feyman diagrams are disconnected. According to a basic
result of QFT, all Green functions can be written
in terms of connected Green functions, i.e.
Green functions that are the sum of connected Feynman
diagrams.
In this section, we show that this elimination can be carried
out by a purely algebraic method, using the fact that
$\calSC$ can be equipped with a second coproduct.

\subsection{A second coproduct on $\calSC$}
For the calculation of connected diagrams, it is 
convenient to define a second coproduct $\delta$
on the symmetric algebra $\calSC$ by
$\Deltad a =1 \otimes a + a \otimes 1$ for 
$a$ in $\calC$, extended to $\calSC$ by
algebra morphism.  Besides, the counit $\counitd$ is defined by
$\counitd(1)=1$, $\counitd(a)=0$ for $a$ in $\calC$ and extended by
algebra morphism.
We denote by $\calSdC$ the resulting bialgebra.
This is the standard symmetric algebra on 
$\calC$ considered as a vector space.
The Sweedler notation for the action of $\Deltad$
on an element $u$ of $\calSC$
is $\Deltad u = \sum u\is1\otimes u\is2$.

We prove now the following crucial
\begin{proposition}
$\calSdC$ is a right $\calSC$-comodule coalgebra
for the right coaction $\beta=\Delta$.
\end{proposition}
\begin{proof}
The map $\beta : \calSdC\rightarrow \calSdC\otimes \calSC$
defined by $\beta=\Delta$ is a right coaction 
of $\calSC$ on $\calSdC$ because we obviously have
(see ref.\cite{Majid} p.22 or ref.\cite{Klimyk} p.29)
$(\beta\otimes\Id)\beta=(\Id\otimes\Delta)\beta$
and $(\Id\otimes\counit)\beta = \Id$.
Thus, $\calSdC$ is a right $\calSC$-comodule coalgebra
if we can prove the two properties
(see ref.\cite{Majid} p.23 or ref.\cite{Klimyk} p.352)
\begin{eqnarray}
(\counitd\otimes\Id)\beta &=& 1\counitd, 
\label{counitdabst}
\\
(\Deltad\otimes\Id)\beta &=&
(\Id\otimes\Id\otimes\mu)(\Id\otimes\tau\otimes\Id)
(\beta\otimes\beta)\Deltad,
\label{DeltaDeltadabst}
\end{eqnarray}
where $\mu$ is the algebra product of $\calSC$ and $\tau$ is the flip.
In Sweedler's notation this becomes
\begin{eqnarray}
\sum \counitd(u\ix1)u\ix2 &=& \counitd(u)1,
\label{counitd}
\\
\sum u\ix1\is1\otimes u\ix1\is2 \otimes u\ix2
 &=&
\sum u\is1\ix1\otimes u\is2\ix1 \otimes u\is1\ix2 u\is2\ix2.
\label{DeltaDeltad}
\end{eqnarray}
We first show equation \eqref{counitd}.
Take $u\in \calS^k(\calC)$ with $k>0$. An example
of such a $u$ in the scalar field algebra is
$u=\varphi^{n_1}(x_1)\dots\varphi^{n_k}(x_k)$.
Then $\counitd(u)=0$ by definition
of $\counitd$. Moreover, $\beta u =\Delta u=\sum u\ix1\otimes u\ix2$,
where all $u\ix1$ and $u\ix2$ belong to 
$\calS^k(\calC)$ by definition of the coproduct $\Delta$.
Thus, $\counitd(u\ix1)=0$ and
$\sum \counitd(u\ix1)u\ix2 = 0 = \counitd(u)1$.
The remaining case is $u\in \calS^0(\calC)$, so that
$u=\lambda 1$ for some complex number $\lambda$.
Then $\sum \counitd(u\ix1)u\ix2 = \lambda \counitd(1) 1= \counitd(u)1$ 
and equation \eqref{counitd} is proved.

To show equation (\ref{DeltaDeltad}) we use a recursive proof.
It is obviously true for $u=1$, assume that
it is true for all elements of 
$\calS^k(\calC)$ up to $k=n$.
Take $u\in\calS^n(\calC)$  and $a\in\calC$.
From $\Deltad(au)=(\Deltad a)(\Deltad u)=\sum a u\is1\otimes u\is2 +
u\is1\otimes au\is2$ we can calculate
\begin{eqnarray*}
(\Deltad\otimes\Id)\Delta(au)
&=&
\sum \Deltad (a\ix1 u\ix1)\otimes a\ix2 u\ix2 
\\&=&
\sum (a\ix1 u\ix1\is1\otimes u\ix1\is2 
+
u\ix1\is1\otimes a\ix1 u\ix1\is2 )\otimes a\ix2 u\ix2
\\&=&
\sum (a\ix1 u\is1\ix1\otimes u\is2\ix1 
+
u\is1\ix1\otimes a\ix1 u\is2\ix1 )\otimes a\ix2 u\is1\ix2 u\is2\ix2
\\&=&
(\Id\otimes\Id\otimes\mu)(\Id\otimes\tau\otimes\Id)
(\Delta\otimes\Delta)\Deltad(au),
\end{eqnarray*}
where we go from the first line to the second
with the expression for $\Deltad(a\ix1 u\ix1)$,
from the second line to the third with the recursion hypothesis
and from the third to the fourth with the expression for
$\Deltad(au)$.
Thus, by linearity, equation (\ref{DeltaDeltad}) is
true for all elements of $\calS^{n+1}(\calC)$.
\end{proof}
For the coalgebra of the scalar field, one might be tempted
to replace all $\varphi^0(x_i)$ by 1, the unit of $\calSC$.
However, if we do this the coproduct of an element $u$ will
contain the terms $u\otimes 1$ and $1\otimes u$ that
spoil the validity of eq.~(\ref{counitd}).

\subsection{The connected chronological product}
\label{ccpsect}
We denote the reduced coproduct
$\Deltadu u = \Deltad u - 1 \otimes u - u \otimes 1$
with Sweedler's notation
$\Deltadu u = \sum u\isu1 \otimes u\isu2$.
\begin{definition}\label{defTc}
For $u \in \calSC$ with $\counitd(u)=0$, we define the
connected chronological product $\Tc(u)$ as
 \begin{eqnarray*}
       \Tc(u) &=& -\sum_{n=1}^\infty \frac{(-1)^n}{n}
       T(u\isu1) {\dots} T(u\iisu{n}).
 \end{eqnarray*}
\end{definition}
For notational convenience, we sometimes omit from now on
the sum sign corresponding to the coproduct:
we write 
$T(u\isu1) {\dots} T(u\iisu{n})$ for
$\sum T(u\isu1) {\dots} T(u\iisu{n})$.
This is called the enhanced Sweedler notation.
Note that the connected chronological product is related
to the Eulerian idempotent \cite{Sweedler}
for $u\in\ker\counitd$
\begin{eqnarray*}
   e(u) &=&  -\sum_{n=1}^\infty \frac{(-1)^n}{n}
      u\isu1\,\dots\, u\iisu{n},
\end{eqnarray*}
in the sense that the operator $T$ is applied to
each term $u\iisu{i}$. Recall that the first Eulerian
idempotent projects onto the primitive elements of a
connected cocommutative bialgebra.
Reciprocally, we can express $T$ in terms of $\Tc$ by
\begin{lemma}\label{TdeTc}
For $u\in\ker\counitd$
\begin{eqnarray*}
      T(u) &=& 
       \sum_{n=1}^\infty \frac{1}{n!} \Tc(u\isu1)\dots \Tc(u\iisu{n}).
\end{eqnarray*}
\end{lemma}
\begin{proof}
From the definition of $\Tc$, we have
\begin{eqnarray*}
\sum_{n=1}^\infty \frac{1}{n!} \Tc(u\isu1)\dots \Tc(u\iisu{n})
&=&
\sum_{n=1}^\infty \frac{(-1)^n}{n!} 
\sum_{i_1,\dots,i_n}  \frac{(-1)^{i_1+\dots+i_n}}
    {i_1\dots i_n}
T(u\isu1)\dots T(u\iisu{i_1+\dots+i_n}),\\
&=& \sum_{k=1}^\infty T(u\isu1)\dots T(u\iisu{k})
\sum_{n=1}^k \frac{(-1)^{k+n}}{n!} 
\sum_{i_1+\dots+i_n=k}  \frac{1}{i_1\dots i_n}.
\end{eqnarray*}
The sum over $n$ and $i_1,\dots,i_n$
is the coefficient of $x^k$ in the series
expansion of $\ee^{\log(1+x)}$. From
$\ee^{\log(1+x)}=1+x$ we deduce that the sum is $\delta_{k,1}$.
Therefore
\begin{eqnarray*}
\sum_{n=1}^\infty \frac{1}{n!} \Tc(u\isu1)\dots \Tc(u\iisu{n})
&=& T(u\isu1) =T(u).
\end{eqnarray*}
\end{proof}
For example,
\begin{eqnarray*}
T(a) &=& \Tc(a),\\
T(ab) &=& \Tc(ab)+\Tc(a)\Tc(b),\\
T(abc) &=& \Tc(abc)+\Tc(ab)\Tc(c)+\Tc(a)\Tc(bc)+\Tc(ac)\Tc(b)+
\Tc(a)\Tc(b)\Tc(c),
\end{eqnarray*}
where $a$, $b$ and $c$ are elements of $\calC$.
The relations between $T$ and $\Tc$ were given
by Haag \cite{Haag58}, Epstein and
Glaser \cite{Epstein}, Brunetti and Fredenhagen \cite{Brunetti},
D\"utsch and Fredenhagen \cite{Dutsch}
and Mestre and Oeckl \cite{Mestre}.

The chronological product $T(u)$ is a coregular action:
$T(u)=\sum t(u\ix1)u\ix2$, where the coproduct is $\Delta$.
The connected chronological product $\Tc(u)$ is defined
in terms of $T(u)$ through the coproduct $\Deltad$.
Therefore, it is rather surprising that $\Tc$
is also a coregular action: there is an element $\tc$
of $\calSC^*$ such that $\Tc(u)=\sum \tc(u\ix1)u\ix2$.
As we shall see, this is a consequence of the fact
that $\calSdC$ is a comodule coalgebra over $\calSC$.

\begin{proposition}
The connected chronological product is a coregular action:
$\Tc(u)=\sum \tc(u\ix1)u\ix2$
for $u\in\ker\counit$, with
 \begin{eqnarray*}
       \tc(u) &=& -\sum_{n=1}^\infty \frac{(-1)^n}{n}
       t(u\isu1) {\dots} t(u\iisu{n}).
 \end{eqnarray*}
\end{proposition}
\begin{proof}
From $\Deltad u = \Deltadu u + 1\otimes u + u \otimes 1$,
it is straightforward to show that equation
\eqref{DeltaDeltadabst}
implies
\begin{eqnarray*}
(\Id\otimes\Id\otimes\mu)(\Id\otimes\tau\otimes\Id)
(\beta\otimes\beta)\Deltadu &=&
(\Deltadu\otimes\Id)\beta.
\end{eqnarray*}
In Sweedler's notation,
\begin{eqnarray}
 \sum u\isu1\ix1\otimes u\isu2\ix1 \otimes u\isu1\ix2 u\isu2\ix2
    &=&
  \sum u\ix1\isu1\otimes u\ix1\isu2 \otimes u\ix2.
\label{coregu}
\end{eqnarray}
Take now two coregular actions
$A(u)=\sum a(u\ix1) u\ix2$ and $B(u)=\sum b(u\ix1) u\ix2$. We have,
using equation \eqref{coregu} for $u\in\ker\counit$,
\begin{eqnarray*} 
  \sum A(u\isu1) B( u\isu2) &=& \sum 
    a(u\isu1\ix1) b(u\isu2\ix1) u\isu1\ix2 u\isu2\ix2
    \\&=&
    \sum a(u\ix1\isu1) b(u\ix1\isu2) u\ix2
  =\sum c(u\ix1) u\ix2,
\end{eqnarray*}
with $c(u)=\sum a(u\isu1) b(u\isu2)$.
Therefore, $\sum A(u\isu1) B( u\isu2)$ is a coregular action.
Using this argument recursively, we obtain that,
if $A_1(u)=\sum a_1(u\ix1) u\ix2,\dots,A_k(u)=\sum a_k(u\ix1) u\ix2$
are $k$ coregular actions, then
$\sum A_1(u\isu1)\dots A_k( u\iisu{k})$ is a
coregular action $\sum c(u\ix1) u\ix2$, with
$c(u)=\sum a_1(u\isu1)\dots a_k( u\iisu{k})$.
This proves that all terms of $\Tc$ are coregular
actions, so that their sum is also a coregular action.
\end{proof}

\subsection{The linked-cluster theorem}
The name of the connected chronological product comes from the
fact that, for the coalgebra of the scalar field, 
$\tc(u)$ is made of exactly the connected diagrams
of $t(u)$. This was proved, for example by Mestre and Oeckl
\cite{Mestre}.
We sketch an alternative proof of this. For a S-matrix
$S=T\big(\exp(\lambda a)\big)$, with $a\in\calC$, we can calculate 
an expression relating $t(\ee^{\lambda a})$ and 
$\tc(\ee^{\lambda a})$.
First, we have for $k>0$ and $n>0$,
\begin{eqnarray}
\Deltadu^{k-1} a^n &=&
\sum_{i_1+\dots+i_k=n} \frac{n!}{i_1!\dots i_k!}
a^{i_1}\otimes\dots\otimes a^{i_k},
\label{deltasan}
\end{eqnarray}
where all $i_j$ are strictly positive integers.
Therefore,
\begin{eqnarray*}
t(a^n) &=&
\sum_{k=1}^n \frac{1}{k!} 
\sum_{i_1+\dots+i_k=n} \frac{n!}{i_1!\dots i_k!}
\tc(a^{i_1})\dots \tc(a^{i_k}).
\end{eqnarray*}
If we write $E=\exp(\lambda a)$, this gives us
\begin{eqnarray*}
t(E) &=& 1+\sum_{n=1}^\infty \frac{\lambda^n}{n!}
\sum_{k=1}^n \frac{1}{k!} 
\sum_{i_1+\dots+i_k=n} \frac{n!}{i_1!\dots i_k!}
\tc(a^{i_1})\dots \tc(a^{i_k})\\
&=& 1+ \sum_{k=1}^\infty \frac{1}{k!}
\sum_{n=k}^\infty 
\sum_{i_1+\dots+i_k=n} \frac{1}{i_1!\dots i_k!}
\tc((\lambda a)^{i_1})\dots \tc((\lambda a)^{i_k})\\
&=& 1+ \sum_{k=1}^\infty \frac{1}{k!} {\big(\tc(E-1)\big)}^k
= \ee^{\tc(E-1)}.
\end{eqnarray*}
The last line was obtained because all $i_j>0$.
This result can also be obtained from the identity
$\Deltadu (E-1)=(E-1)\otimes (E-1)$.
The fact that $\tc(E-1)$ contains only connected
Feynman diagrams  follows from the fact that
the logarithm of $t(E)$ is the sum of
all connected vacuum diagrams \cite{Itzykson}.

The same proof holds for the $T$-products, so that
\begin{eqnarray*}
S &=& T(E) = \exp\big(\Tc(E-1)\big).
\end{eqnarray*}
If we define a connected $S$-matrix
by $S_c=\Tc(E-1)$, we obtain the
linked-cluster theorem \cite{LindgrenMorrison}
$S=\ee^{S_c}$.

\subsection{A noncommutative analogue}
If the tensor bialgebra $\calTC$ is used instead of
the symmetric bialgebra $\calSC$, the construction is
similar.
We start from the bialgebra $\calTC$ and we define
the coalgebra $\calTdC$ to be the vector space $\calTC$
endowed with the deconcatenation coproduct:
$\Deltad 1= 1\otimes 1$,
$\Deltad a= a\otimes 1 + 1 \otimes a$ for $a\in\calC$ and
\begin{eqnarray*}
\Deltad u &=& u\otimes 1 + 1 \otimes u
  +\sum_{k=1}^{n-1} a_1\dots a_k  \otimes a_{k+1}\dots a_n,
\end{eqnarray*}
for $u=a_1\dots a_n\in \calT^n(\calC)$ and $n>1$.
The counit $\counitd$ of $\calTdC$ is defined by
$\counitd(1)=1$ and $\counitd(u)=0$ if $u\in \calT^n(\calC)$
with $n>0$.
If $\calTdC$ is equipped with the concatenation product,
then $\calTdC$ is not a bialgebra because the coproduct is
not an algebra morphism.
Loday and Ronco \cite{LodayRonco4} showed that
the deconcatenation coproduct and the concatenation
product satisfy the compatibility rule
$\Deltad (uv) =
(u\otimes 1)\Deltad v + \Deltad (u) (1\otimes v)
-u\otimes v$, which makes
$\calTdC$ a unital infinitesimal bialgebra.
Note that, if $u=a$, the compatibility rule becomes
\begin{eqnarray}
\Deltad (av) &=& (a\otimes 1)\Deltad v + 1 \otimes av,
\label{Deltasaunc}
\end{eqnarray}
for $a\in\calC$ and $v\in \calTdC$.

We have the following
\begin{proposition}
$\calTdC$ is a right $\calTC$-comodule coalgebra
for the right coaction $\beta=\Delta$.
\end{proposition}
\begin{proof}
The proof of condition \eqref{counitd} on the counit
is exactly the same as for the symmetric case.
We prove \eqref{DeltaDeltad} recursively.
It is obviously true for $u=1$, assume that
this is true for elements of degree up to $n$.
Take $u$ an element of degree $n$ and $a\in\calC$.
We rewrite equation \eqref{Deltasaunc} as
$\Deltad (au) = \sum a u\is1\otimes u\is2 + 1\otimes au$.
Thus,
\begin{eqnarray*}
(\Delta\otimes\Delta)\Deltad au &=&
\sum a\ix1 u\is1\ix1\otimes a\ix2 u\is1\ix2\otimes u\is2\ix1 \otimes
u\is2\ix2 + 
\sum 1\otimes 1\otimes a\ix1 u\ix1\otimes a\ix2 u\ix2.
\end{eqnarray*}
\begin{eqnarray}
(\Id\otimes\tau\otimes\Id)
(\Delta\otimes\Delta)\Deltad au &=&
\sum a\ix1 u\is1\ix1\otimes u\is2\ix1 \otimes a\ix2 u\is1\ix2\otimes
u\is2\ix2 
\nonumber\\&&+ \sum 1\otimes a\ix1 u\ix1\otimes 1 \otimes a\ix2 u\ix2.
\label{ttoott}
\end{eqnarray}
Thus,
\begin{eqnarray*}
(\Deltad\otimes\Id)\Delta (au) &=&
\sum \Deltad (a\ix1 u\ix1)\otimes a\ix2 u\ix2
\\ &=&
\sum a\ix1 u\ix1\is1\otimes u\ix1\is2 \otimes a\ix2 u\ix2
+
\sum 1\otimes a\ix1 u\ix1\otimes a\ix2 u\ix2
\\ &=&
\sum a\ix1 u\is1\ix1\otimes u\is2\ix1 \otimes a\ix2 u\is1\ix2 u\is2\ix2 + 
\sum 1\otimes a\ix1 u\ix1\otimes  a\ix2 u\ix2
\\&=&
(\Id\otimes\Id\otimes\mu)(\Id\otimes\tau\otimes\Id)
(\Delta\otimes\Delta)\Deltad (au),
\end{eqnarray*}
where we go from the first line to the second using
equation \eqref{Deltasaunc}, from the second to the third
with the recursion
hypothesis and from the third to the fourth
using equation \eqref{ttoott}.
This completes the proof.
\end{proof}
Inspired by the analogue of the first Eulerian idempotent
defined by Loday and Ronco \cite{LodayRonco4} for 
connected unital infinitesimal bialgebras,  
(for $u\in\ker\counitd$)
\begin{eqnarray*}
e &=& -\sum_{n=1}^\infty (-1)^n u\isu1\,\dots\, u\iisu{n},
\end{eqnarray*}
we define the connected chronological product $\Tc$ by
\begin{eqnarray*}
\Tc(u) &=& -\sum_{n=1}^\infty (-1)^n T(u\isu1) {\dots} T(u\iisu{n}),
\end{eqnarray*}
or, reciprocally,
\begin{eqnarray*}
T(u) &=& \sum_{n=1}^\infty \Tc(u\isu1) {\dots} \Tc(u\iisu{n}),
\end{eqnarray*}
still for $u\in\ker\counitd$.
Again, $\Tc$ is a coregular action if $T$ is a coregular action.

\section{Renormalization}

Renormalization is a fundamental aspect of quantum field
theory. It was discovered because the values of many Feynman diagrams
are divergent.
After several attempts, notably by Dyson \cite{Dyson},
the problem was essentially solved by Bogoliubov
\cite{Bogoliubov}.
The renormalization theory found in most textbooks
\cite{Itzykson} is a development of the Bogoliubov
approach called the BPHZ renormalization.
However, it appeared recently that the original
Bogoliubov approach has decisive advantage over
the BPHZ renormalization. In particular, it
can be used for the renormalization of
quantum field theory in curved spacetime
\cite{Brunetti2,Hollands4}.

We first present Bogoliubov's solution in Hopf algebraic
terms, then we consider in more detail a simplified model.

\subsection{The Bogoliubov formula}

Bogoliubov (\cite{Bogoliubov}, section 26.2)
and Epstein-Glaser \cite{Epstein} showed that the
relation between the bare (i.e. divergent)
chronological product $T$ and the renormalized
chronological product $T'$ is
\begin{eqnarray}
T'(u) &=& \sum_{n=1}^\infty \frac{1}{n!}
  T\big(\calO(u\isu1)\dots \calO(u\iisu{n})\big),
\label{TpdeT}
\end{eqnarray}
for $u\in\ker\counit$
and $T'(1)=T(1)=1$.
In equation (\ref{TpdeT}),
$\calO$ is a linear operator $\ker\counitd \rightarrow \calC$
called a generalized vertex \cite{Bogoliubov}.
Epstein and Glaser proved that the standard BPHZ renormalization
is a consequence of this formula \cite{Epstein}.
Note that the renormalized chronological product $T'$
is not in general a coregular action.

To see the effect of the operator $\calO$ we calculate
$T'(E)$ for $E=\ee^{\lambda a}$, with $a\in\calC$. We first use equation
(\ref{deltasan}) to write
\begin{eqnarray*}
T'(a^n) &=&
\sum_{k=1}^n \frac{1}{k!} 
\sum_{i_1+\dots+i_k=n} \frac{n!}{i_1!\dots i_k!}
T\big(\calO(a^{i_1})\dots \calO(a^{i_k})\big),
\end{eqnarray*}
where all $i_j>0$.
This gives us
\begin{eqnarray*}
T'(E) &=& 1 + \sum_{n=1}^\infty \frac{\lambda^n}{n!} T'(a^n) 
=  1+\sum_{n=1}^\infty
\sum_{k=1}^n \frac{1}{k!} 
\sum_{i_1+\dots+i_k=n} \frac{1}{i_1!\dots i_k!}
T\big(\calO((\lambda a)^{i_1})\dots \calO((\lambda a)^{i_k})\big)
\\&=& 1+
\sum_{k=1}^\infty \frac{1}{k!} 
\sum_{i_1,\dots,i_k} \frac{1}{i_1!\dots i_k!}
T\big(\calO((\lambda a)^{i_1})\dots \calO((\lambda a)^{i_k})\big)
\\&=&
1+\sum_{k=1}^n \frac{1}{k!} 
T(\calO(\ee^{\lambda a}-1)\dots \calO(\ee^{\lambda a}-1))
=
T\big(\exp(\calO(\ee^{\lambda a}-1))\big).
\end{eqnarray*}

If we define $a'\in \calC$ by
\begin{eqnarray}
a' &=& \frac{1}{\lambda}\calO(\ee^{\lambda a}-1) = 
      \calO(a) +\sum_{n=2}^\infty \frac{\lambda^{n-1}}{n!} 
       \calO(a^n),
\label{apdea}
\end{eqnarray}
the previous equality can be rewritten
$T'(\ee^{\lambda a})=T(\ee^{\lambda a'})$.
In other words, the change of chronological product
from $T$ to $T'$ amounts to a change of
Lagrangian from $a$ to $a'$.
This result was obtained by Hollands and Wald
\cite{Hollands4} who showed that it holds also in
curved spacetime.

In flat spacetime, the chronological product satisfies
$T(a)=T'(a)=a$ for $a\in\calC$. Therefore,
$a=T'(a)=T(\calO(a))=\calO(a)$ because
$\calO(a)\in\calC$. This implies $\calO(a)=a$
and the renormalized Lagrangian starts with the
unrenormalized one. The terms with $n>1$ in equation
(\ref{apdea}) are called the renormalization \emph{counterterms}.
In curved spacetime the situation is more complicated and
Hollands and Wald \cite{Hollands} showed that we have
in general $T'(a)=\sum t'(a\ix1) T(a\ix2)$, where
$t'$ is a linear map from $\calC$ to $\mathbb{C}$.
In that case $\calO(a)= \sum t'(a\ix1) a\ix2$.

\subsection{The renormalization group: preparation}
In this section, we define a product on
linear maps $\Lambda : \calSVp \rightarrow V$
for any vector space $V$ on the complex numbers, where
$\calSV= \mathbb{C} 1 \oplus \calSVp$ is the symmetric
Hopf algebra on $V$, with coproduct $\Deltad$
and counit $\counitd$. The Sweedler notation for
the coproduct $\Deltad$ is again
$\Deltad u = \sum u\is1 \otimes u\is2$.

\begin{definition}\label{defLambdan}
If $\LSV$ denotes the set of linear maps from 
$\calSV$ to $\calSV$, the \emph{convolution product}
of two elements $f$ and $g$ of $\LSV$ is the element
$f*g$ of $\LSV$ defined by
$(f*g)(u)=\sum f(u\is1) g(u\is2)$, where
$u\in \calSV$.
The \emph{convolution powers} of an element
$f$ of $\LSV$ are the elements $f^{*n}$ of $\LSV$
defined by $f^{*0}=\counitd 1$ and
$f^{*n}=f*f^{*(n-1)}$ for any integer $n>0$.

In particular, if we denote by $\LSVV$ the
set of linear maps from $\calSVp$ to $V$,
we first extend $\Lambda$ to $\calSV$ by
$\Lambda(1)=0$. Then, we define the convolution
powers $\Lambda^{*n}$ as above and the 
\emph{convolution exponential} $\ee^{*\Lambda}$ by
\begin{eqnarray*}
\ee^{*\Lambda}(u) &=& \sum_{n=0}^\infty \frac{1}{n!} \Lambda^{*n}(u).
\end{eqnarray*}
\end{definition}
Note that the exponential is well defined (i.e. the sum is finite)
because $\Lambda(1)=0$ implies that, for $u\in  \calS^k(V)$, 
$\Lambda^{*n}(u)=0$ for $n>k$.
The following special cases are illustrative:
$\ee^{*\Lambda}(1)=1$,
$\ee^{*\Lambda}(a)=\Lambda(a)$,
$\ee^{*\Lambda}(ab)=\Lambda(ab)+\Lambda(a)\Lambda(b)$ 
and 
\begin{eqnarray*}
\ee^{*\Lambda}(abc) &=& \Lambda(abc)+\Lambda(a)\Lambda(bc)
+\Lambda(b)\Lambda(ac) +\Lambda(c)\Lambda(ab)
+\Lambda(a)\Lambda(b)\Lambda(c),
\end{eqnarray*}
for $a$, $b$ and $c$ in $V$.
Note also that $\ee^{*\Lambda}$
maps $\calSVp$ to $\calSVp$.  We first prove the useful lemma
\begin{lemma}\label{DeltadeeLambdau}
For $\Lambda\in\LSVV$ and $u\in\calSV$, we have
$\Deltad \big(\ee^{*\Lambda}(u)\big) = 
\sum
\ee^{*\Lambda}(u\is1) \otimes \ee^{*\Lambda}(u\is2)$.
\end{lemma}
\begin{proof}
The space $\LSV$ equipped with the convolution product is a
commutative algebra with unit $\bfun=\counitd 1$.
We denote by $\calA$ the subalgebra generated by $\LSVV$ (where the
elements $\Lambda$ of $\LSVV$ are extended to $\calSV$
by $\Lambda(1)=0$). For any $\Lambda$ of $\LSVV$,
$\Lambda(u)$ is an element of $V$. Thus, it is primitive
and
\begin{eqnarray*}
\Deltad \Lambda(u) &=& \Lambda(u) \otimes 1 + 1 \otimes \Lambda(u)
=(\Lambda\otimes\counitd 1 + \counitd 1\otimes \Lambda)\Deltad u
=(\Lambda\otimes\bfun + \bfun\otimes \Lambda)\Deltad u.
\end{eqnarray*}
Therefore, it is natural to equip $\calA$ with
the structure of a Hopf algebra by defining
the coproduct
$\Delta \Lambda=\Lambda\otimes\bfun + \bfun\otimes \Lambda$
for $\Lambda\in\LSVV$ and extending it to $\calA$
by algebra morphism.
The equality $\Delta \ee^{*\Lambda}=\ee^{*\Lambda}\otimes
\ee^{*\Lambda}$ follows from the fact that 
the exponential of a primitive element is group-like.
The lemma is a consequence of the fact that $\calSV$
is a $\calA$-module coalgebra for the action
$\Lambda\triangleright u=\Lambda(u)$.
\end{proof}

A second lemma will be useful to derive recursive proofs.
\begin{lemma} \label{eeLambdaau}
For $\Lambda\in\LSVV$, $a\in V$ and $u\in\calSV$, we have
$\ee^{*\Lambda}(au) = 
    \sum \Lambda(a u\is1) \ee^{*\Lambda} (u\is2)$.
\end{lemma}
\begin{proof}
We first show recursively that, for
$a\in V$, $u\in\calSV$ and $n>0$, then
\begin{eqnarray}
\Lambda^{*n}(au) &=& n \sum\Lambda(au\is1) \Lambda^{*(n-1)}(u\is2).
\label{Lambdan(au)}
\end{eqnarray}
This is true for $n=1$ because
$\Lambda(au)=\sum \Lambda(au\is1) \counit(u\is2)
=\sum \Lambda^{*1}(au\is1) \Lambda^{*0}(u\is2)$.
Assume that equation \eqref{Lambdan(au)} is true up to $n$.
Then,
\begin{eqnarray*}
\Lambda^{*(n+1)}(au) &=& \sum \Lambda\big((au)\is1\big) 
  \Lambda^{*n}\big((au)\is2\big)
=\sum \Lambda(au\is1) \Lambda^{*n}(u\is2)+ 
   \sum \Lambda(u\is1) \Lambda^{*n}(au\is2)
\\&=&
\sum \Lambda(au\is1) \Lambda^{*n}(u\is2)+ n\sum \Lambda(u\is1)  
  \Lambda(au\is2)\Lambda^{*(n-1)}(u\is3)
\\&=&
(n+1)\sum \Lambda(au\is1) \Lambda^{*n}(u\is2),
\end{eqnarray*}
where we used the coassociativity and cocommutativity of
the coproduct and the commutativity of the product.
The lemma follows because
\begin{eqnarray*}
\ee^{*\Lambda}(au) &=& \counitd(au)+\sum_{n=1}^\infty \frac{1}{n!} 
   \Lambda^{*n}(au)
=\sum_{n=1}^\infty \frac{1}{(n-1)!} \sum
\Lambda(au\is1)\Lambda^{*(n-1)}(u\is2)
\\&=&
\sum \Lambda(au\is1)\ee^{*\Lambda}(u\is2),
\end{eqnarray*}
where we used the fact that $\counitd(au)=\counitd(a)\counitd(u)=0$
because $\counitd(a)=0$.
\end{proof}

We are now ready to define a product on $\LSVV$ by 
\begin{definition}
If $\Lambda'$ and $\Lambda$ are in $\LSVV$,
the product of $\Lambda'$ and $\Lambda$ is the 
element $\Lambda' \bullet \Lambda$ of $\LSVV$ defined by
\begin{eqnarray*}
(\Lambda' \bullet \Lambda)(u) &=& \Lambda'\big(\ee^{*\Lambda}(u)\big).
\end{eqnarray*}
\end{definition}

This definition enables us to write the last lemma
of this section. 
\begin{lemma}\label{eeLamLamp}
For $\Lambda'$ and $\Lambda$ in $\LSVV$ and $u\in \calSV$, we have
$\ee^{*\Lambda'}\big(\ee^{*\Lambda}(u)\big)=
\ee^{*(\Lambda'\bullet\Lambda)}(u)$.
\end{lemma}
\begin{proof}
The lemma is true for $u=1$ and $u=a\in V$ because
$\ee^{*\Lambda'}\big(\ee^{*\Lambda}(1)\big)=\ee^{*\Lambda'}(1)=1
=\ee^{*(\Lambda'\bullet\Lambda)}(1)$
and
$\ee^{*\Lambda'}\big(\ee^{*\Lambda}(a)\big)=\ee^{*\Lambda'}(\Lambda(a))=
\Lambda'(\Lambda(a))=(\Lambda'\bullet\Lambda)(a)=
\ee^{*(\Lambda'\bullet\Lambda)}(a)$.
Assume that the lemma is true for all elements of 
$\calS^k(V)$ up to $k=n$. Take $a\in V$, $u\in\calS^n(V)$ and use lemma
\ref{eeLambdaau} to calculate
\begin{eqnarray*}
\ee^{*\Lambda'}\big(\ee^{*\Lambda}(au)\big) &=&
\sum \ee^{*\Lambda'}\big(\Lambda(a u\is1) \ee^{*\Lambda} (u\is2)\big)
\end{eqnarray*}
If we denote $\Lambda(a u\is1)$ by $a'$ and
$\ee^{*\Lambda} (u\is2)$ by $u'$, we can use
lemma \ref{eeLambdaau} again
\begin{eqnarray*}
\ee^{*\Lambda'}(a'u') &=& \sum \Lambda'(a' u\is1')\ee^{*\Lambda'}(u\is2').
\end{eqnarray*}
Lemma \ref{DeltadeeLambdau} enables us to calculate
$\sum  u\is1'\otimes u\is2' = \delta u'
=\delta \ee^{*\Lambda} (u\is2)=\sum  \ee^{*\Lambda} (u\is2) \otimes
\ee^{*\Lambda} (u\is3)$. Therefore,
\begin{eqnarray*}
\ee^{*\Lambda'}\big(\ee^{*\Lambda}(au)\big) &=&
\sum \Lambda'\big(\Lambda(a u\is1)\ee^{*\Lambda} (u\is2)\big)
  \ee^{*\Lambda'}(\ee^{*\Lambda} (u\is3))
\\&=&
\sum \Lambda'\big(\Lambda(a u\is1)\ee^{*\Lambda} (u\is2)\big)
  \ee^{*(\Lambda'\bullet\Lambda)} (u\is3),
\end{eqnarray*}
where we used the recursion hypothesis to evaluate 
$\ee^{*\Lambda'}(\ee^{*\Lambda} (u\is3))$.
Lemma \ref{eeLambdaau} and the definition of $\Lambda' \bullet \Lambda$
yield
\begin{eqnarray*}
\ee^{*\Lambda'}\big(\ee^{*\Lambda}(au)\big) &=&
\sum \Lambda'\big(\ee^{*\Lambda} (a u\is1)\big)
  \ee^{*(\Lambda'\bullet\Lambda)} (u\is2)
=
\sum (\Lambda'\bullet\Lambda) (a u\is1) \ee^{*(\Lambda'\bullet\Lambda)} (u\is2)
\\&=&
\ee^{*(\Lambda'\bullet\Lambda)} (au),
\end{eqnarray*}
where we used lemma \ref{eeLambdaau} again to conclude.
Thus, the lemma is true for $au \in \calS^{n+1}(V)$.
\end{proof}
These lemmas lead us to the main result of this section,
\begin{proposition} The vector space
$\LSVV$ endowed with the product $\bullet$ is a unital associative
algebra. 
The unit of this algebra is the map $\Lambda_0$
such that $\Lambda_0(a)=a$ for $a\in V$ and
$\Lambda_0(u)=0$ for $u\in \calS^n(V)$ with $n>1$.
The invertible elements of this algebra are exactly
the $\Lambda$ such that the restriction of $\Lambda$ to $V$
is invertible as a linear map from $V$ to $V$.
\end{proposition}
\begin{proof}
Associativity follows essentially from lemma
\ref{eeLamLamp} because, for $\Lambda_1$,
$\Lambda_2$ and $\Lambda_3$ in $\LSVV$ we have
\begin{eqnarray*}
\big((\Lambda_1 \bullet \Lambda_2) \bullet \Lambda_3 \big)(u) 
    &=&
 (\Lambda_1 \bullet \Lambda_2)\big(\ee^{*\Lambda_3}(u) \big)
=
 \Lambda_1 \big( \ee^{*\Lambda_2}(\ee^{*\Lambda_3}(u)) \big)
=
 \Lambda_1 \big( \ee^{*(\Lambda_2\bullet\Lambda_3)}(u) \big)
\\&=&
\big(\Lambda_1 \bullet (\Lambda_2 \bullet \Lambda_3 )\big)(u).
\end{eqnarray*}

$\Lambda_0$ is the unit of the algebra because
$\ee^{*\Lambda_0}(u)=u$ for any $u\in\calSV$. This is
true for $u=1$ by the definition of $\ee^{*\Lambda_0}$
and for $u=a\in V$ by the definition of $\Lambda_0$.
Assume that this is true for $u$ of degree up to $n$.
Take $u$ of degree $n$ and use
lemma \ref{eeLambdaau}:
$\ee^{*\Lambda_0}(au)=\sum 
\Lambda_0 (a u\is1) \ee^{*\Lambda_0} (u\is2)$.
$\Lambda_0 (a u\is1)=0$ if the degree of $u\is1$ is larger
than 0. Thus $\ee^{*\Lambda_0}(au)=\sum 
\Lambda_0 (a) \ee^{*\Lambda_0} (u) = au$ by the recursion hypothesis.
Thus $(\Lambda\bullet\Lambda_0)(u)=\Lambda\big(\ee^{*\Lambda_0}(u)\big)
=\Lambda(u)$. Similarly,
$(\Lambda_0\bullet\Lambda)(u)=\Lambda_0\big(\ee^{*\Lambda}(u)\big)$.
$\Lambda_0$ is the identity on the elements of degree 1 and 0 on
the elements of degree different from 1. The element of degree
1 of $\ee^{*\Lambda}(u)$ is $\Lambda(u)$, thus
$(\Lambda_0\bullet\Lambda)(u)=\Lambda(u)$.

To prove the invertibility property, consider an invertible element
$\Lambda$ with inverse $\Lambda^{-1}$. Then
$\Lambda^{-1}\bullet\Lambda=\Lambda_0$ and, on any element $a$ of $V$,
$(\Lambda^{-1}\bullet\Lambda)(a)=\Lambda_0(a)=a$. Thus,
$\Lambda^{-1}\big(\Lambda(a)\big)=a$ and $\Lambda$ is invertible
as a map from $V$ to $V$.
Reciprocally, take a  $\Lambda$ invertible
as a map from $V$ to $V$ with inverse $\Lambda'$. We shall
construct recursively the inverse $\Lambda^{-1}$ of $\Lambda$
in the algebra $\LSVV$. For $a\in V$ we have
$\Lambda^{-1}(a)=\Lambda'(a)$. To see how the recurrence works,
we calculate the next term.  For $a$ and $b$ in $V$ we are
looking for a $\Lambda^{-1}$ such that
$(\Lambda^{-1}\bullet\Lambda)(ab)=\Lambda_0(ab)=0$. Thus,
$0=\Lambda^{-1}(\ee^{*\Lambda}(ab))
=\Lambda^{-1}(\Lambda(ab)+\Lambda(a)\Lambda(b))$. 
This defines $\Lambda^{-1}(\Lambda(a)\Lambda(b))=
-\Lambda'(\Lambda(ab))$, because $\Lambda^{-1}$ is
$\Lambda'$ on $V$. The map $\Lambda$ being bijective on $V$,
this defines $\Lambda^{-1}$ on $\calS^2(V)$.
Assume now that $\Lambda^{-1}$ is defined on $\calS^k(V)$
for all $k<n$, take $a_1,\dots,a_n$ in $V$ and
put $u=a_1,\dots,a_n$ in $\calS^n(V)$.
We want to solve $(\Lambda^{-1}\bullet\Lambda)(u)=0$,
with $(\Lambda^{-1}\bullet\Lambda)(u)=\Lambda^{-1}(\ee^{*\Lambda}(u))$.
The term of highest degree in 
$\ee^{*\Lambda}(u)$ is $\Lambda(a_1)\dots\Lambda(a_n)$.
The inverse $\Lambda^{-1}$ is defined on all the other terms,
thus the equation $(\Lambda^{-1}\bullet\Lambda)(u)=0$ defines
$\Lambda^{-1}$ on  $\Lambda(a_1)\dots\Lambda(a_n)$.
In other words, $\Lambda^{-1}$ is now uniquely 
defined on $\calS^n(V)$. Therefore, $\Lambda^{-1}$ is uniquely 
defined on $\calSVp$.
\end{proof}

\subsection{Renormalization group: QFT}
If $\calO$ is a linear map from $\calSCp$ to $\calC$,
we saw in equation (\ref{apdea}) that the renormalization
encoded in $\calO$ can be considered as a change of
Lagrangian from $a$ to $a'$ with
\begin{eqnarray*}
a' &=& \calO(a) +\sum_{n=2}^\infty
\frac{\lambda^{n-1}}{n!} 
       \calO(a^n).
\end{eqnarray*}
Thus, it is possible to consider $a'$ as the result of the action
of $\Lambda$ on $a$: $a' = \calO\triangleright a$.
If we renormalize the Lagrangian $a'$ with the renormalization
encoded in a map $\calO'$, we obtain a new Lagrangian
$a''= \calO'\triangleright a'= \calO'\triangleright (\calO\triangleright a)$. 
The first terms of $a''$ are
\begin{eqnarray*}
 a'' &=& \calO'(\calO(a)) + \frac{\lambda}{2}
 \big(\calO'(\calO(a^2)) + 
                  \calO'(\calO(a)\calO(a))\big)
\\&& + \frac{\lambda^2}{6} \big(\calO'(\calO(a^3)) + 
                  3\calO'(\calO(a)\calO(a^2))+
                  \calO'(\calO(a)\calO(a)\calO(a))\big)
   + O(\lambda^3)
\end{eqnarray*}
The main result of this section is
\begin{proposition}
If $\calO$ and $\calO'$ are in $\LSVV$
and $a\in \calC$, then
$\calO'\triangleright (\calO\triangleright a)=(\calO'\bullet\calO)
\triangleright a$.
\end{proposition}
\begin{proof}
Equation \eqref{deltasan} for $\Deltadu^{k-1} a^n$ gives us
\begin{eqnarray*}
\ee^{*\Lambda}(a^n) &=&  \sum_{k=1}^\infty \frac{1}{k!}
\sum_{i_1+\dots+i_k=n} \frac{n!}{i_1!\dots i_k!}
\calO(a^{i_1})\dots\calO(a^{i_k}),
\end{eqnarray*}
where all the $i_j$ are strictly positive integers. Thus, for 
$E=\ee^{\lambda a}$,
\begin{eqnarray*}
\ee^{*\Lambda}(\frac{E-1}{\lambda}) &=& \sum_{n=1}^\infty \lambda^{n-1}
     \sum_{k=1}^\infty \frac{1}{k!} \sum_{i_1+\dots+i_k=n}
\frac{\calO(a^{i_1})}{i_1!}\dots\frac{\calO(a^{i_k})}{i_k!}
\\&=&
 \sum_{k=1}^\infty \frac{1}{k!} \sum_{i_1,\dots,i_k} \lambda^{i_1+\dots+i_k-1}
\frac{\calO(a^{i_1})}{i_1!}\dots\frac{\calO(a^{i_k})}{i_k!}.
\end{eqnarray*}
On the other hand, 
\begin{eqnarray*}
\Lambda'\triangleright a' &=& \Lambda'\Big(\sum_{k=1}^\infty \frac{\lambda^{k-1}}{k!}
    (a')^k\Big),
\end{eqnarray*}
with $a'=\sum_{i=1}^\infty \lambda^{i-1} \Lambda(a^i)/i!$. Therefore,
\begin{eqnarray*}
\Lambda'\triangleright a' &=& \Lambda'\Big(\sum_{k=1}^\infty \frac{1}{k!}
    \sum_{i_1,\dots,i_k} \lambda^{i_1+\dots+i_k-1}
     \frac{\calO(a^{i_1})}{i_1!}\dots\frac{\calO(a^{i_k})}{i_k!}\Big).
\end{eqnarray*}
Thus,
\begin{eqnarray*}
\Lambda'\triangleright a' &=& \Lambda'\Big(\ee^{*\Lambda}
    \big(\frac{E-1}{\lambda}\big)\Big)
          = (\Lambda'\bullet\Lambda)\big(\frac{E-1}{\lambda}\big)
          = (\Lambda'\bullet\Lambda)\triangleright a,
\end{eqnarray*}
where we used lemma \ref{eeLamLamp} and the fact that, for any $\calO$,
$\calO\triangleright a = (1/\lambda)\calO(E-1)$.
\end{proof}
In standard QFT, the linear maps $\calO$ satisfy 
$\calO(a)=a$. Thus, they are invertible for the 
product $\bullet$ and they form a group, which is one of the many
faces of the \emph{renormalization group}.

\subsection{Connected renormalization}
We argued that, in QFT, the connected chronological product
is physically more useful than the standard chronological product.
Thus, it is important to investigate how connected
chronological products are renormalized.

\begin{proposition}
The relation between the connected renormalized 
chronological product $\Tc'$ and the connected
chronological product $\Tc$ is, for $u\in\ker\counitd$,
\begin{eqnarray*}
\Tc'(u) &=& \sum_{n=1}^\infty \frac{1}{n!}
  \Tc\big(\calO(u\isu1)\dots \calO(u\iisu{n})\big),
\end{eqnarray*}
\end{proposition}
\begin{proof}
We first note that we can use definition
\ref{defLambdan} to rewrite equation \eqref{TpdeT}
under the form $T'(u)=T(\ee^{*\Lambda}(u))$. 
Lemma \ref{TdeTc} expresses $T(\ee^{*\Lambda}(u))$ 
in terms of connected chronological products.
To evaluate this expression we need 
$\Deltadu^{n-1} \ee^{*\Lambda}(u)$.
The identity $\Deltad v = v \otimes 1 + 1 \otimes v
+ \Deltadu v$ transforms lemma \ref{DeltadeeLambdau}
into
$\Deltadu \ee^{*\Lambda}(u) = \big(\ee^{*\Lambda}\otimes \ee^{*\Lambda}\big)
  \Deltadu u$.
By iterating and using the coassociativity of $\Deltadu$ we find
\begin{eqnarray}
\Deltadu^{n-1} \ee^{*\Lambda}(u) &=& \sum
\ee^{*\Lambda}(u\isu1)\otimes \dots \otimes \ee^{*\Lambda}(u\iisu{n}),
\label{deltaunee}
\end{eqnarray}
and we can rewrite the renormalized chronological product
in terms of the bare connected chronological products as
\begin{eqnarray*}
T'(u) &=& 
\sum_{n=1}^\infty \frac{1}{n!}
   \Tc(\ee^{*\Lambda}(u\isu1)) \dots \Tc(\ee^{*\Lambda}(u\iisu{n})).
\end{eqnarray*}
To conclude, we use definition \ref{defTc} to
express $\Tc'(u)$ in terms of $T'$. Then, we expand 
each $T'(u\iisu{i})$ using the last equation
\begin{eqnarray*}
\Tc'(u) &=&
\sum_{n=1}^\infty \frac{(-1)^{n+1}}{n} T'(u\isu1)\dots T'(u\iisu{n})
\\&=&
\sum_{n=1}^\infty \frac{(-1)^{n+1}}{n} 
\sum_{i_1,\dots,i_n}  \frac{1}{i_1!\dots i_n!}
\Tc(\ee^{*\Lambda}(u\isu1))\dots \Tc(\ee^{*\Lambda}(u\iisu{i_1+\dots+i_n}))\\
&=& \sum_{k=1}^\infty \Tc(\ee^{*\Lambda}(u\isu1))\dots
\Tc(\ee^{*\Lambda}(u\iisu{k}))
\sum_{n=1}^k \frac{(-1)^{n+1}}{n} 
\sum_{i_1+\dots+i_n=k}  \frac{1}{i_1!\dots i_n!}.
\end{eqnarray*}
The sum over $n$ and $i_1,\dots,i_n$
is the coefficient of $x^k$ in the series
expansion of $\log(\ee^x)=\log(1+(\ee^x-1))$. From
$\log(\ee^x)=x$ we deduce that the sum is $\delta_{k,1}$.
Therefore
$\Tc'(u)=\Tc(\ee^{*\Lambda}(u))$ and the lemma is proved.
\end{proof}
In other words, the connected chronological product is
renormalized with the same formula and the same
generalized vertices $\calO$ as the standard chronological
product. Such an expression for the renormalization
of the connected chronological product was used, for
instance, by Hollands \cite{Hollands6}.

\subsection{A simplified model}\label{simpmodsect}
In QFT, the linear maps $\calO$ have a very particular form.
In the example of the $\varphi^4(x)$ theory, the
Lagrangian is $a=\int_{\mathbb{R}^4} \varphi^4(x) \dd x$ and
\begin{eqnarray*}
    \calO(a^n) &=& 
           C_1^{(n)} \int_{\mathbb{R}^4} \varphi^4(x) \dd x+
          C_3^{(n)} \int_{\mathbb{R}^4} \varphi^2(x) \dd x 
         +
          C_2^{(n)} \int_{\mathbb{R}^4} \varphi(x)
             \big(\partial\cdot\partial - m^2\big)\varphi(x) \dd x,
\end{eqnarray*}
where $C_1^{(n)}$, $C_2^{(n)}$ and $C_2^{(n)}$ are real numbers
related to the charge, wavefunction and mass renormalization
\cite{Bogoliubov}.
Such a Lagrangian cannot be manipulated directly with our approach
because the integral over $\mathbb{R}^4$ involves an infinite
number of points. However, as explained in section \ref{chronsect},
it can be given a meaning in the perturbative approach.
A more serious problem is the presence of derivatives in
$ \int_{\mathbb{R}^4} \varphi(x)
             \big(\partial\cdot\partial - m^2\big)\varphi(x) \dd x$.
To deal with such terms, we must include derivatives of fields 
into our algebra $\calC$. This poses several problems that are
debated by Stora, Boas, D\"utsch and Fredenhagen
\cite{Dutsch02,Dutsch03,Dutsch04,Dutsch06},
concerning the status of the Action Ward Identity or
whether the fields should be taken on-shell or off-shell.
Before the situation is fully clarified, we can propose
a model without derivatives (i.e. where the divergences
are only logarithmic, in the QFT parlance \cite{Itzykson}).
In that case, it was shown in a paper with Bill Schmitt
\cite{BrouderSchmitt}, that the coalgebra $\calC$ has
to be replaced by a bialgebra $\calB$ and that renormalization
becomes a functor on bialgebras.
In the case of the scalar field, the product is defined by
$\varphi^n(x_i)\cdot \varphi^m(x_j)=\delta_{ij} \varphi^{n+m}(x_i)$.
It can be checked that, with this product, the coalgebra of the scalar
field becomes indeed a commutative bialgebra $\calB$.

This simplified model can be extended to 
any commutative and cocommutative 
bialgebra $\calB$ by defining the maps $\calO$ as
$\calO(u)=\sum \lambda(u\ix1) \prod u\ix2$,
where $\lambda$ is a linear map from $\calSBp$ to $\mathbb{C}$
and $\prod u$ is defined as follows: if 
$u=a\in \calB$, then $\prod u=a$, if
$u=a_1\dots a_n \in \calS^n(\calB)$, then 
$\prod u=a_1\cdot{\dots} \cdot a_n$ where the product
$\cdot$ is in $\calB$. With this definition, it is clear
that $\calO$ is a linear map from $\ker \counitd$ to $\calB$.
The fact that such $\calO$ form a subgroup of the renormalization
group defined in the previous paragraph follows from the
existence of a bialgebraic structure on 
$\calS(\calSBp)$, studied in detail in \cite{BrouderSchmitt}.
If the product of $n$ elements $u_1,\dots,u_n$
of $\calSBp$ in $\calS^n(\calSBp)$ is denoted by
$u_1\vee\dots\vee u_n$, the renormalization coproduct
$\Delta_R$ defined in \cite{BrouderSchmitt} can be
written, for $u\in \calSBp$,
\begin{eqnarray*}
\Delta_R u &=& \sum_{n=1}^\infty \frac{1}{n!}
  \sum u\isu1\ix1 \vee \dots\vee  u\iisu{n}\ix1 \otimes
  (\prod u\isu2\ix2) \dots  (\prod u\iisu{n}\ix2).
\end{eqnarray*}

This construction has a functorial noncommutative analogue
\cite{BrouderSchmitt}. 

\subsection{The renormalization group: the noncommutative case}
In this section, we want to describe the renormalization group
in the noncommutative case. We first define a product on
linear maps $\Lambda : \calTVp \rightarrow V$
for any vector space $V$ on the complex numbers, where
$\calTV= \mathbb{C} 1 \oplus \calTVp$ is the 
tensor algebra on $V$, with deconcatenation coproduct $\Deltad$
and counit $\counitd$. Let us denoted by $\LTVV$ the
set of linear maps from $\calTVp$ to $V$.

For $\Lambda$ in $\LTVV$, we
first define the noncommutative analogue of $\ee^{*\Lambda}$,
that we denote by $\eei{\Lambda}$.
\begin{definition}
Let $\Lambda\in\LTVV$, we define the convolution powers
$\Lambda^{*n}$ as in definition \ref{defLambdan},
with the symmetric coproduct replaced by the 
deconcatenation coproduct.
Moreover, we define the linear map
$\eei{\Lambda} : \calTV \rightarrow \calTV$ by
\begin{eqnarray*}
\eei{\Lambda}(u) &=& \sum_{n=0}^\infty \Lambda^{*n}(u),
\end{eqnarray*}
for $u\in\calTV$.
\end{definition}
Note that $\eei{\Lambda}$ is well defined because, for
$u\in  \calT^k(V)$, $\Lambda^{*n}(u)=0$ if $n>k$.
The following special cases are illustrative:
$\eei{\Lambda}(1)=1$,
$\eei{\Lambda}(a)=\Lambda(a)$,
$\eei{\Lambda}(ab)=\Lambda(ab)+\Lambda(a)\Lambda(b)$ 
and 
\begin{eqnarray*}
\eei{\Lambda}(abc) &=& \Lambda(abc)+\Lambda(a)\Lambda(bc)
+\Lambda(ab)\Lambda(c)
+\Lambda(a)\Lambda(b)\Lambda(c)
\end{eqnarray*}
for $a$, $b$ and $c$ in $V$.
Note also that $\eei{\Lambda}$
maps $\calTVp$ to $\calTVp$.
As for the commutative case, we have the useful lemma
\begin{lemma}\label{DeltadeeLambdaunc}
For $\Lambda\in\LTVV$ and $u\in\calTV$, we have
$\Deltad \eei{\Lambda}(u) = \big(\eei{\Lambda}\otimes \eei{\Lambda}\big)
  \Deltad u$.
\end{lemma}
\begin{proof}
We give a detailed proof because
unital infinitesimal algebra are not as well studied as
Hopf algebras.  We first show recursively the identity
\begin{eqnarray}
\Deltad \Lambda^{*n}(u) &=&
  \sum_{k=0}^n \sum \Lambda^{*k}(u\is1 ) \otimes
          \Lambda^{*(n-k)}(u\is2 ).
\label{DeltadLambdannc}
\end{eqnarray}
For $n=0$, equation \eqref{DeltadLambdannc} is satisfied because
\begin{eqnarray*}
\Deltad \Lambda^{*0}(u) &=& \counitd(u) 1\otimes 1
= \sum \counitd(u\is1) 1\otimes \counitd(u\is2) 1
=  \sum \Lambda^{*0}(u\is1) \otimes \Lambda^{*0}(u\is2).
\end{eqnarray*}
Equation \eqref{DeltadLambdannc} is 
obviously true for $u=1$ and all $n>0$.
Thus, we take from now on $u\in\calTVp$.
Assume that equation \eqref{DeltadLambdannc}  is true up to $n$, then
the definition of $\Lambda^{(n+1)}(u)$, the relation 
$\Deltad (au) = (a\otimes 1)\Deltad u + 1 \otimes au$,
and the recursion hypothesis imply
\begin{eqnarray*}
\Deltad \Lambda^{*(n+1)}(u) &=&
\sum_{k=0}^n \sum \Lambda(u\is1)\Lambda^{*k}(u\is2)\otimes
  \Lambda^{*(n-k)}(u\is3)
+\sum 1 \otimes \Lambda(u\is1)\Lambda^{*n}(u\is2)
\\&=&
\sum_{k=0}^n \sum \Lambda^{*(k+1)}(u\is1)\otimes
  \Lambda^{*(n-k)}(u\is2)
+ 1 \otimes \Lambda^{*(n+1)}(u)
\\&=&
\sum_{k=0}^{n+1} \sum \Lambda^{*k}(u\is1)\otimes
  \Lambda^{*(n+1-k)}(u\is2).
\end{eqnarray*}
The lemma follows by summing both sides of equation
\eqref{DeltadLambdannc} over $n$.
\end{proof}

A second lemma is very close to its commutative analogue.
\begin{lemma} \label{eeLambdaaunc}
For $\Lambda\in\LTVV$, $a\in V$ and $u\in\calSV$ we have
$\eei{\Lambda}(au) = 
    \sum \Lambda(a u\is1) \eei{\Lambda} (u\is2)$.
\end{lemma}
\begin{proof}
We first show that, if
$a\in V$, $u\in\calSVp$ and $n>0$, then
\begin{eqnarray}
\Lambda^{*n}(au) &=& \sum \Lambda(a u\is1) \Lambda^{*(n-1)}(u\is2).
\label{Lambdan(au)nc}
\end{eqnarray}
For any $n>0$, the definition of the convolution powers and the 
compatibility rule
$\Deltad (au) = (a\otimes 1)\Deltad u
    + 1 \otimes au$
imply
\begin{eqnarray*}
\Lambda^{*n}(au) &=& 
  \sum\Lambda((au)\is1) \Lambda^{*(n-1)}((au)\is2)
\\&=&
  \sum\Lambda(au\is1) \Lambda^{*(n-1)}(u\is2)
  + \sum\Lambda(1) \Lambda^{*(n-1)}(au)
=
  \sum\Lambda(au\is1) \Lambda^{*(n-1)}(u\is2),
\end{eqnarray*}
because $\Lambda(1)=0$.
Thus,
\begin{eqnarray*}
 \eei{\Lambda} (au) &=& \counitd(au)1
  + \sum_{n=1}^\infty \Lambda^{*n}(au)
=
  \sum_{n=1}^\infty \sum \Lambda(a u\is1) \Lambda^{*(n-1)}(u\is2)
  =\sum \Lambda(a u\is1)\eei{\Lambda} (u\is2),
\end{eqnarray*}
where we used $\counitd(au)=\counitd(a)\counitd(u)=0$,
since $\counitd(a)=0$.
\end{proof}

We are now ready to define a product on $\LTVV$ by 
\begin{definition}
If $\Lambda'$ and $\Lambda$ are in $\LTVV$,
the product of $\Lambda'$ and $\Lambda$ is the
element $\Lambda' \bullet \Lambda$ of $\LTVV$ defined by
\begin{eqnarray*}
(\Lambda' \bullet \Lambda)(u) &=& \Lambda'\big(\eei{\Lambda}(u)\big).
\end{eqnarray*}
\end{definition}

This definition enables us to write the last lemma
of this section. 
\begin{lemma}\label{eeLamLampnc}
For $\Lambda'$ and $\Lambda$ in $\LTVV$ and $u\in \calTV$, we have
$\eei{\Lambda'}\big(\eei{\Lambda}(u)\big)=\eei{\Lambda'\bullet\Lambda}(u)$.
\end{lemma}
\begin{proof}
The proof is the same as for the commutative case.
\end{proof}
We can now state the main result of this section,
\begin{proposition}The vector space
$\LTVV$ endowed with the product $\bullet$ is a unital associative
algebra. 
The unit of this algebra is the map $\Lambda_0$
such that $\Lambda_0(a)=a$ for $a\in V$ and
$\Lambda_0(u)=0$ for $u\in \calT^n(V)$ with $n>1$.
The invertible elements of this algebra are exactly
the $\Lambda$ such that the restriction of $\Lambda$ to $V$
is invertible as a linear map from $V$ to $V$.
In particular, the subset of $\LTVV$ characterized
by $\Lambda(a)=a$ for $a\in V$ is a group.
\end{proposition}
\begin{proof}
The proof is the same as for the commutative case.
\end{proof}

\section{Conclusion}
This paper described the first steps of a complete
description of QFT in Hopf algebraic terms.
Although these steps look encouraging, many open
problems still have to be solved.
The main one is analytical: the use of a finite
number of points is not really satisfactory and
we should allow for coalgebras containing elements
such as $\int \varphi^n(x) g(x) \dd x$ for some
test functions $g$.

Other open problems are easier. We list now three of them:
(i) The renormalization approach presented here is equivalent to
the Connes-Kreimer approach because both are equivalent to
the standard BPHZ renormalization \cite{BrouderSchmitt}.
However, it would be quite interesting to describe this
equivalence explicitly.
(ii) We proved that a QFT is renormalized once its connected
chronological product is renormalized. In fact, a deeper
result is true: a QFT is renormalized once its one-particle
irreducible diagrams are renormalized \cite{Itzykson}. 
To cast this result into our framework, we would need
to write the connected chronological product $\Tc$
in terms of a one-particle irreducible chronological
product. Although such a connection was announced by
Epstein and Glaser \cite{Epstein}, it was described
as complicated and was apparently never published.
Three possible solutions to this problem have been 
explored~\cite{Stora,BrouderBSL,BFP}
Similarly, it would be worthwhile to determine a
Hopf algebraic expression of Green functions in 
terms of $n$-particle irreducible functions, which is usually
done by Legendre transformation techniques \cite{Pismak4}. 
(iii) It would be important to develop the analogue of the
constructions presented in this paper to the case
of gauge theories. Along that line, van Suijlekom
obtained the remarkable result that the
Ward and Slavnov-Taylor identities generate a Hopf
ideal of the Hopf algebra of renormalization
\cite{Suijlekom}. It would be nice to see how
this result can be adapted to our framework.

The most original aspect of this work is the 
determination of noncommutative analogues 
of some QFT concepts (i.e. the replacement of
$\calSC$ by $\calTC$, or of $\calSV$ by $\calTV$).
Such a noncommutative analogue was first determined for
quantum electrodynamics \cite{BFI} and it lead to the
definition of a noncommutative Fa\`a di Bruno
algebra \cite{BFKI}, generalized to many variables
by Anshelevich et al. \cite{Anshelevich}.
These algebras provide an effective way to manipulate
series in noncommutative variables.
The noncommutative constructions defined in
the present paper can also be useful for
that purpose.

The present approach enables us to
recover the Feynman diagram formulation of QFT,
but its most interesting aspect is that it is 
defined at the operator level. For example,
in our notation, the relation between connected
and standard chronological product is given not
only at the level of the coregular actions
$t$ and $\tc$, but at the level of the
maps from $\calSC$ to $\calSC$ (i.e. the
relation between $T$ and $\Tc$). 
As a consequence, we can calculate Green functions
such as 
\begin{eqnarray*}
G_\rho(x_1,\dots,x_n) &=& 
\frac{\rho\Big(T\big(\varphi(x_1)\dots\varphi(x_n) 
       \ee^{\lambda a}\big)\Big)}
{\rho\Big(T\big(\ee^{\lambda a}\big)\Big)},
\end{eqnarray*}
where $\rho$ is a map from $\calSC$ to $\mathbb{C}$.
Such more general Green functions are the basic objects
of the quantum field theory with initial correlations
(or QFT of degenerate systems) which is well suited to
the calculation of highly-correlated systems
\cite{BrouderEuroLett}.

\begin{acknowledgement}
I am very indebted to Raymond Stora for
sending me some of his fascinating manuscripts
and for pointing out the Hopf algebraic structure used
in Ruelle's work \cite{Ruelle}.
I am grateful to Jean-Yves Thibon for pointing out
Burge's article \cite{Burge}.
Exciting discussions with Alessandra Frabetti,
Jose Maria Gracia-Bondia
and William Schmitt are gratefully acknowledged.
Finally, I am very grateful to Prof. Zeidler,
whose encouragement provided a crucial motivation
to write this paper.
This work was partly supported by the ANR grant
ANR-05-JCJC-0044-01.
\end{acknowledgement}


\end{document}